\documentclass[aps,prd,twocolumn,superscriptaddress,showpacs]{revtex4} 

\usepackage[dvips]{epsfig,color}
\usepackage{graphicx}
\usepackage{wrapfig,rotating}
\usepackage{amssymb,amsmath,array}

\begin{document}


\newcommand{\fix}[1]{} 
\newcommand{\MET}{\mbox{$\raisebox{.3ex}{$\not\!$}E_T$}}
\newcommand{\METVEC}{\mbox{$\raisebox{.3ex}{$\not\!$}{\vec E}_T$}}
\newcommand{\MPTVEC}{\mbox{$\raisebox{.3ex}{$\not\!$}{\vec {p}_T}$}}

\def\DZero{D0~}

\newcommand{\Et}{\mbox{$E_T$}}
\newcommand{\et}{\mbox{$E_T$}}
\newcommand{\Pt}{\mbox{$p_T$}}
\newcommand{\pt}{\mbox{$p_T$}}
\newcommand{\Ht}{\mbox{$H_T$}}
\newcommand{\met}{\mbox{$\protect \raisebox{0.3ex}{$\not$}\Et$}}
\newcommand{\mpt}{\mbox{$\protect \raisebox{0.3ex}{$\not$}\Pt$}}



\title{Measurement of the {\boldmath WW+WZ} Production Cross Section Using a Matrix Element Technique in Lepton + Jets Events}


\affiliation{Institute of Physics, Academia Sinica, Taipei, Taiwan 11529, Republic of China} 
\affiliation{Argonne National Laboratory, Argonne, Illinois 60439, USA} 
\affiliation{University of Athens, 157 71 Athens, Greece} 
\affiliation{Institut de Fisica d'Altes Energies, Universitat Autonoma de Barcelona, E-08193, Bellaterra (Barcelona), Spain} 
\affiliation{Baylor University, Waco, Texas 76798, USA} 
\affiliation{Istituto Nazionale di Fisica Nucleare Bologna, $^{bb}$University of Bologna, I-40127 Bologna, Italy} 
\affiliation{Brandeis University, Waltham, Massachusetts 02254, USA} 
\affiliation{University of California, Davis, Davis, California 95616, USA} 
\affiliation{University of California, Los Angeles, Los Angeles, California 90024, USA} 
\affiliation{Instituto de Fisica de Cantabria, CSIC-University of Cantabria, 39005 Santander, Spain} 
\affiliation{Carnegie Mellon University, Pittsburgh, Pennsylvania 15213, USA} 
\affiliation{Enrico Fermi Institute, University of Chicago, Chicago, Illinois 60637, USA}
\affiliation{Comenius University, 842 48 Bratislava, Slovakia; Institute of Experimental Physics, 040 01 Kosice, Slovakia} 
\affiliation{Joint Institute for Nuclear Research, RU-141980 Dubna, Russia} 
\affiliation{Duke University, Durham, North Carolina 27708, USA} 
\affiliation{Fermi National Accelerator Laboratory, Batavia, Illinois 60510, USA} 
\affiliation{University of Florida, Gainesville, Florida 32611, USA} 
\affiliation{Laboratori Nazionali di Frascati, Istituto Nazionale di Fisica Nucleare, I-00044 Frascati, Italy} 
\affiliation{University of Geneva, CH-1211 Geneva 4, Switzerland} 
\affiliation{Glasgow University, Glasgow G12 8QQ, United Kingdom} 
\affiliation{Harvard University, Cambridge, Massachusetts 02138, USA} 
\affiliation{Division of High Energy Physics, Department of Physics, University of Helsinki and Helsinki Institute of Physics, FIN-00014, Helsinki, Finland} 
\affiliation{University of Illinois, Urbana, Illinois 61801, USA} 
\affiliation{The Johns Hopkins University, Baltimore, Maryland 21218, USA} 
\affiliation{Institut f\"{u}r Experimentelle Kernphysik, Karlsruhe Institute of Technology, D-76131 Karlsruhe, Germany} 
\affiliation{Center for High Energy Physics: Kyungpook National University, Daegu 702-701, Korea; Seoul National University, Seoul 151-742, Korea; Sungkyunkwan University, Suwon 440-746, Korea; Korea Institute of Science and Technology Information, Daejeon 305-806, Korea; Chonnam National University, Gwangju 500-757, Korea; Chonbuk National University, Jeonju 561-756, Korea} 
\affiliation{Ernest Orlando Lawrence Berkeley National Laboratory, Berkeley, California 94720, USA} 
\affiliation{University of Liverpool, Liverpool L69 7ZE, United Kingdom} 
\affiliation{University College London, London WC1E 6BT, United Kingdom} 
\affiliation{Centro de Investigaciones Energeticas Medioambientales y Tecnologicas, E-28040 Madrid, Spain} 
\affiliation{Massachusetts Institute of Technology, Cambridge, Massachusetts 02139, USA} 
\affiliation{Institute of Particle Physics: McGill University, Montr\'{e}al, Qu\'{e}bec, Canada H3A~2T8; Simon Fraser University, Burnaby, British Columbia, Canada V5A~1S6; University of Toronto, Toronto, Ontario, Canada M5S~1A7; and TRIUMF, Vancouver, British Columbia, Canada V6T~2A3} 
\affiliation{University of Michigan, Ann Arbor, Michigan 48109, USA} 
\affiliation{Michigan State University, East Lansing, Michigan 48824, USA}
\affiliation{Institution for Theoretical and Experimental Physics, ITEP, Moscow 117259, Russia}
\affiliation{University of New Mexico, Albuquerque, New Mexico 87131, USA} 
\affiliation{Northwestern University, Evanston, Illinois 60208, USA} 
\affiliation{The Ohio State University, Columbus, Ohio 43210, USA} 
\affiliation{Okayama University, Okayama 700-8530, Japan} 
\affiliation{Osaka City University, Osaka 588, Japan} 
\affiliation{University of Oxford, Oxford OX1 3RH, United Kingdom} 
\affiliation{Istituto Nazionale di Fisica Nucleare, Sezione di Padova-Trento, $^{cc}$University of Padova, I-35131 Padova, Italy} 
\affiliation{LPNHE, Universite Pierre et Marie Curie/IN2P3-CNRS, UMR7585, Paris, F-75252 France} 
\affiliation{University of Pennsylvania, Philadelphia, Pennsylvania 19104, USA}
\affiliation{Istituto Nazionale di Fisica Nucleare Pisa, $^{dd}$University of Pisa, $^{ee}$University of Siena and $^{ff}$Scuola Normale Superiore, I-56127 Pisa, Italy} 
\affiliation{University of Pittsburgh, Pittsburgh, Pennsylvania 15260, USA} 
\affiliation{Purdue University, West Lafayette, Indiana 47907, USA} 
\affiliation{University of Rochester, Rochester, New York 14627, USA} 
\affiliation{The Rockefeller University, New York, New York 10065, USA} 
\affiliation{Istituto Nazionale di Fisica Nucleare, Sezione di Roma 1, $^{gg}$Sapienza Universit\`{a} di Roma, I-00185 Roma, Italy} 

\affiliation{Rutgers University, Piscataway, New Jersey 08855, USA} 
\affiliation{Texas A\&M University, College Station, Texas 77843, USA} 
\affiliation{Istituto Nazionale di Fisica Nucleare Trieste/Udine, I-34100 Trieste, $^{hh}$University of Trieste/Udine, I-33100 Udine, Italy} 
\affiliation{University of Tsukuba, Tsukuba, Ibaraki 305, Japan} 
\affiliation{Tufts University, Medford, Massachusetts 02155, USA} 
\affiliation{Waseda University, Tokyo 169, Japan} 
\affiliation{Wayne State University, Detroit, Michigan 48201, USA} 
\affiliation{University of Wisconsin, Madison, Wisconsin 53706, USA} 
\affiliation{Yale University, New Haven, Connecticut 06520, USA} 
\author{T.~Aaltonen}
\affiliation{Division of High Energy Physics, Department of Physics, University of Helsinki and Helsinki Institute of Physics, FIN-00014, Helsinki, Finland}
\author{B.~\'{A}lvarez~Gonz\'{a}lez$^v$}
\affiliation{Instituto de Fisica de Cantabria, CSIC-University of Cantabria, 39005 Santander, Spain}
\author{S.~Amerio}
\affiliation{Istituto Nazionale di Fisica Nucleare, Sezione di Padova-Trento, $^{cc}$University of Padova, I-35131 Padova, Italy} 

\author{D.~Amidei}
\affiliation{University of Michigan, Ann Arbor, Michigan 48109, USA}
\author{A.~Anastassov}
\affiliation{Northwestern University, Evanston, Illinois 60208, USA}
\author{A.~Annovi}
\affiliation{Laboratori Nazionali di Frascati, Istituto Nazionale di Fisica Nucleare, I-00044 Frascati, Italy}
\author{J.~Antos}
\affiliation{Comenius University, 842 48 Bratislava, Slovakia; Institute of Experimental Physics, 040 01 Kosice, Slovakia}
\author{G.~Apollinari}
\affiliation{Fermi National Accelerator Laboratory, Batavia, Illinois 60510, USA}
\author{J.A.~Appel}
\affiliation{Fermi National Accelerator Laboratory, Batavia, Illinois 60510, USA}
\author{A.~Apresyan}
\affiliation{Purdue University, West Lafayette, Indiana 47907, USA}
% ...
\collaboration{CDF Collaboration\footnote{With visitors from $^a$University of Massachusetts Amherst, Amherst, Massachusetts 01003,
$^b$Istituto Nazionale di Fisica Nucleare, Sezione di Cagliari, 09042 Monserrato (Cagliari), Italy,
$^c$University of California Irvine, Irvine, CA  92697, 
$^d$University of California Santa Barbara, Santa Barbara, CA 93106
$^e$University of California Santa Cruz, Santa Cruz, CA  95064,
$^f$CERN,CH-1211 Geneva, Switzerland,
$^g$Cornell University, Ithaca, NY  14853, 
$^h$University of Cyprus, Nicosia CY-1678, Cyprus, 
$^i$University College Dublin, Dublin 4, Ireland,
$^j$University of Fukui, Fukui City, Fukui Prefecture, Japan 910-0017,
$^k$Universidad Iberoamericana, Mexico D.F., Mexico,
$^l$Iowa State University, Ames, IA  50011,
$^m$University of Iowa, Iowa City, IA  52242,
$^n$Kinki University, Higashi-Osaka City, Japan 577-8502,
$^o$Kansas State University, Manhattan, KS 66506,
$^p$University of Manchester, Manchester M13 9PL, England,
$^q$Queen Mary, University of London, London, E1 4NS, England,
$^r$Muons, Inc., Batavia, IL 60510,
$^s$Nagasaki Institute of Applied Science, Nagasaki, Japan, 
$^t$National Research Nuclear University, Moscow, Russia,
$^u$University of Notre Dame, Notre Dame, IN 46556,
$^v$Universidad de Oviedo, E-33007 Oviedo, Spain, 
$^w$Texas Tech University, Lubbock, TX  79609, 
$^x$IFIC(CSIC-Universitat de Valencia), 56071 Valencia, Spain,
$^y$Universidad Tecnica Federico Santa Maria, 110v Valparaiso, Chile,
$^z$University of Virginia, Charlottesville, VA  22906,
$^{aa}$Yarmouk University, Irbid 211-63, Jordan,
$^{ii}$On leave from J.~Stefan Institute, Ljubljana, Slovenia, 
}}
\noaffiliation

\today

\begin{abstract}

We present a measurement of the $WW+WZ$ production cross section observed in a
final state consisting of an identified electron or muon, two jets,
and missing transverse energy.  The measurement is carried out in a
data sample corresponding to up to 4.6~fb$^{-1}$ of integrated
luminosity at $\sqrt{s} = 1.96$ TeV collected by the CDF II detector.
Matrix element calculations are used to separate the diboson signal
from the large backgrounds.  The $WW+WZ$ cross section is measured to
be $17.4\pm3.3$~pb, in agreement with standard model predictions.  A fit to the dijet
invariant mass spectrum yields a compatible cross section measurement.

\end{abstract}

\pacs{14.80.Bn, 14.70.Fm, 14.70.Hp, 12.15.Ji}  

\maketitle




\section{Introduction}
\label{sec:intro}

Measurements of the production cross section of pairs of heavy gauge
bosons test the electroweak sector of the standard model (SM).  The
production cross section can be enhanced by anomalous triple gauge
boson interactions~\cite{hagiwara} or from new particles decaying to
pairs of vector bosons.

In this paper, we describe the measurement of the $WW+WZ$ production
cross section in events containing a high-$\pt$ electron or muon and
two hadronic jets.  This event topology is expected when one $W$ boson
in the event decays to an electron or muon and a neutrino, and the
other $W$ or $Z$ boson decays to two quarks ($WW/WZ \rightarrow
\ell\nu qq$).  We consider both the $WW$ and $WZ$ processes as signal because
our limited detector resolution of hadronic jets makes the separation of $W \rightarrow
q\bar{q}'$ from $Z\rightarrow q\bar{q}$ impracticable.

\begin{figure}
\centering \includegraphics[width=0.48\textwidth]{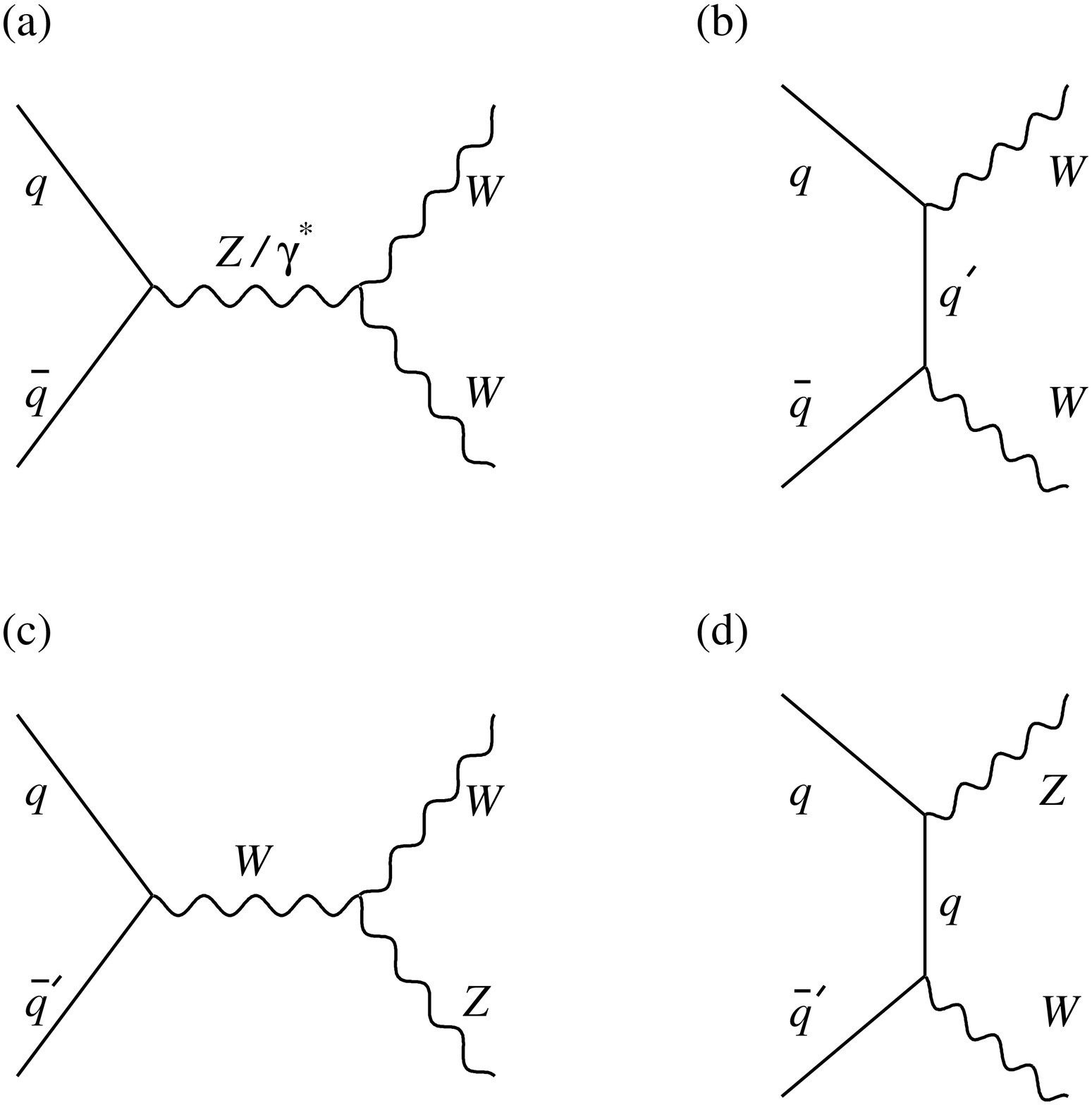}
\caption{\label{fig:ProdDiagrams}Leading-order diagrams for $WW$ $s$-channel (a) and $t$-channel (b) and $WZ$ $s$-channel (c) and $t$-channel (d) production.}
\end{figure}

The leading-order $WW$ and $WZ$ production diagrams are shown in
Fig.~\ref{fig:ProdDiagrams}.  The predicted SM production cross sections
at the Tevatron, calculated at next-to-leading order (NLO), are
$\sigma(p\bar{p}\rightarrow WW) = 11.66\pm0.70$~pb and
$\sigma(p\bar{p} \rightarrow WZ) = 3.46 \pm 0.30$~pb~\cite{VVtheory}.
Both of these production cross sections have been measured previously
at the Tevatron in channels in which both gauge bosons decay
leptonically~\cite{diblepCDF, diblepD0}, and no deviation between
measurement and prediction has been observed.

Hadronic decay modes have higher branching ratios than the leptonic
decays, but the corresponding final states are exposed to large
backgrounds.  The first observation of diboson production at the Tevatron with a
hadronic decay was achieved in events with two jets and large missing
transverse energy at CDF~\cite{metjets}.  Evidence and observation of
the process and decay discussed in this paper, $WW+WZ \rightarrow\ell
\nu qq$, were previously reported by the D0~\cite{d0lvjj} and CDF~\cite{ourPRL} collaborations.
The observation reported by CDF used a matrix element technique
relying on knowledge of the differential cross sections of signal and
background processes to separate signal events from the background. 

The measurement of $WW+WZ \rightarrow \ell\nu qq$ is relevant to the
search for the Higgs boson at the Tevatron.  One of the most powerful
channels used in the search for a Higgs boson with a mass lower than
130 GeV/$c^2$ is the channel in which the Higgs boson is produced in
association with a $W$ boson, with the Higgs boson decaying to a pair
of $b$ quarks and the $W$ boson decaying leptonically ($WH \rightarrow
\ell\nu b\bar{b}$).  A similar matrix element analysis to the one
presented in this paper is employed in the $WH \rightarrow \ell \nu
b\bar{b}$ search at CDF~\cite{WHME}.  A well-established measurement 
of the $WW+WZ$ channel gives us confidence in the similar techniques in the
search for the Higgs boson. Similar issues in background 
modeling and systematic uncertainties are relevant for the two
analyses.  One important difference, however, is that the search for
$WH$ production uses methods to identify jets originating from b-quarks (Ò$b$-taggingÓ), 
whereas the $WW+WZ$ analysis presented in this paper does not use $b$-tagging.

This paper presents the details of the matrix element method used in
the observation of $WW+WZ \rightarrow \ell \nu qq$, but applied to a
larger data sample corresponding to up to 4.6~fb$^{-1}$ of integrated
luminosity taken with the CDF II detector and with some changes in the 
event selection criteria.  In particular, the event selection has been made 
more inclusive so that it resembles that used in the $WH \rightarrow \ell \nu b\bar{b}$
search more closely.

The organization of the rest of this paper is as follows.  Section~\ref{sec:CDF}  describes the apparatus used to carry out the measurement, 
while Section~\ref{sec:sel} describes the event
selection and backgrounds.  The modeling of the signal and background
processes is discussed in Section~\ref{sec:MC}.  Section~\ref{sec:ME}
contains the details of the matrix element technique used for the
measurement.  The systematic uncertainties and results are discussed
in Sections~\ref{sec:Sys} and~\ref{sec:results} respectively.  A fit
to the dijet invariant mass spectrum, performed as a cross check, is
presented in Section~\ref{sec:Mjj}.  Finally, we summarize the
conclusions in Section~\ref{sec:conc}.

\section{CDF II detector}
\label{sec:CDF}

 The CDF II detector is a nearly azimuthally and forward-backward
symmetric detector designed to study $p\bar{p}$ collisions at the
Tevatron.  It is described in detail in Ref.~\cite{CDFdet}.  It
consists of a charged particle tracking system surrounded by
calorimeters and
 muon chambers.  Particle positions and angles are
expressed in a
 cylindric coordinate system, with the $z$ axis along
the proton beam.
 The polar angle, $\theta$, is measured with respect
to the direction
 of the proton beam, and $\phi$ is the azimuthal
angle about the beam
 axis.  The pseudo-rapidity, $\eta$, is defined
as $\eta = - \ln (\tan
\frac {\theta}{2})$.

The momentum of charged particles is measured by the tracking system,
consisting of silicon strip detectors surrounded by an open-cell drift
chamber, all immersed in a 1.4 T solenoidal magnetic field coaxial with the Tevatron beams.  The
silicon tracking system~\cite{SVX} consists of eight layers of silicon
covering the radial region from 1.5 cm to 28 cm from the beam axis.
The drift chamber, or central outer tracker (COT)~\cite{COT}, is
composed of eight superlayers that alternate between axial and 2
degree stereo orientations.  Each superlayer contains 12 sense wires.
The COT covers the radial region from 40 cm to 137 cm and provides
good tracking efficiency for charged particles out to $|\eta|<1.0$.

The tracking system is surrounded by calorimeters which measure the
energies of electrons, photons, and jets of hadronic particles.  The
electromagnetic calorimeters use a scintillating tile and lead
sampling technology, while the hadronic calorimeters are composed
of scintillating tiles with steel absorber.  The calorimeters are
divided into central and plug sections.  The central region, composed
of the central electromagnetic (CEM)~\cite{CEM} and central and
end-wall hadronic calorimeters (CHA and WHA)~\cite{CHAWHA}, covers the
region $|\eta|<1.1$.  The end-plug electromagnetic (PEM)~\cite{PEM}
and end-plug hadronic calorimeters (PHA) extend the coverage to
$|\eta|<3.6$.  The calorimeters have a component called the shower
maximum (ShowerMax)~\cite{ShowerMax} detector located at the depth in
the calorimeter at which the electromagnetic
shower is expected to be widest.  The ShowerMax uses wire chambers and cathode
strips to provide a precise position measurement for electromagnetic
clusters.

A muon system composed of planar multi-wire drift chambers records hits
when charged particles pass through.  Four different sections of the muon
detector are used for the analysis presented here: the central muon
detector (CMU)~\cite{CMU}, the central muon upgrade (CMP), the central
muon extension (CMX), and the barrel muon chambers (BMU).  In the
central region, $|\eta|<0.6$, four layers of chambers located just
outside of the calorimeter make up the CMU system; the CMU is
surrounded by 60 cm of iron shielding and another four layers of
chambers compose the CMP system.  The CMX covers the region with
$0.6<|\eta|<1.0$, while the BMU extends the coverage to $1.0<|\eta|<1.5$.

Cherenkov luminosity counters (CLCs)~\cite{CLC} measure the rate of
inelastic collisions, which can be converted to an instantaneous
luminosity.  The integrated luminosity is calculated from the
instantaneous luminosity measurements.  The CLCs consist of gaseous
Cherenkov counters located at high pseudo-rapidity, 3.6$<|\eta|<$4.6.

The three-level trigger system at CDF is used to reduce the event rate
from 1.7 MHz to about 150 Hz.  The first level uses hardware, while
the second is a mixture of hardware and fast software
algorithms~\cite{XFT}.  The software-based third-level trigger makes use of 
detailed information on the event, very similar to that available offline.

\section{Candidate Event Selection and Backgrounds}
\label{sec:sel}

The event selection can be divided into a baseline selection
corresponding to the topology of our signal, and a variety of vetoes
that are imposed to remove backgrounds.
The baseline selection, the relevant backgrounds, and the vetoes are
all described in more detail below.

A few relevant quantities for the event selection are defined here.
The transverse momentum of a charged particle is $\pt = p \sin
\theta$, where $p$ is the momentum of the charged particle track.  The
analogous quantity measured with calorimeter energies is the
transverse energy, $\et = E \sin \theta$.  The missing transverse
energy, $\METVEC$, is defined by $\METVEC = -\sum_iE_T^i{\hat n}_i$,
where ${\hat n}_i$ is a unit vector perpendicular to the beam axis and
pointing at the $i^{th}$ calorimeter tower.  $\METVEC$ is corrected
for high-energy muons as well as for factors applied to correct
hadronic jet energies.  We define $\MET=|\METVEC|$. Jets are clustered 
using a cone algorithm, with a fixed cone size in which the center of the 
jet is defined as ($\eta^{jet}, \phi^{jet}$) and the size of the jet cone as 
$\Delta R=\sqrt{(\eta^{tower}-\eta^{jet})^2-(\phi^{tower}-\phi^{jet})^2} \leq 0.4$. 

\subsection{Baseline event selection}

\begin{figure}
  \centering \includegraphics[width=0.4\textwidth]{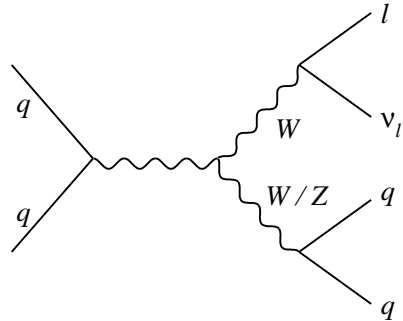}
  \caption{\label{fig:ProdDecay} Decay of $WW/WZ$ events to a charged lepton, neutrino, and two quarks.}
\end{figure}

Figure~\ref{fig:ProdDecay} shows the $WW/WZ$ decay topology
that is considered as the signal in this analysis.  The final state
contains a charged lepton, a neutrino, and two quarks.  We focus on
events in which the charged lepton is an electron or muon.  Events in which
the $W$ boson decays to a $\tau$ lepton may also be considered part of the
signal if a leptonic $\tau$ decay results in an isolated electron or
muon.  The neutrino passes through the detector without depositing
energy; its presence can be inferred in events with $\met$.  The two
quarks will hadronize to form collimated jets of hadrons.  As a
result, our baseline event selection requires events to contain one
high-$\pt$ electron or muon, significant $\met$, and two jets.

Several triggers are used to collect events for this analysis.
Roughly half of the events are selected with a trigger requiring a
high-$\pt$ central electron in the CEM ($\et > 18$ GeV, $|\eta|<1.0$).
Two muon triggers, one requiring hits in both the CMP and CMU and the
other requiring hits in the CMX, collect events with central muons
($\pt > 18$ GeV/$c$, $|\eta|<1.0$).  Finally, a trigger path requiring
large $\met$ and two jets is used to collect events with muons that
were not detected by the central muon triggers.  The $\met$ plus jets
trigger requires $\met > 35$~GeV and two jets with $\et> 10$~GeV.
The jet $\et$ and $\met$ used in the trigger selection are not corrected 
for detector or physics effects.

Further selection criteria are imposed on triggered events offline.
Electron (muon) candidates are required to have $\et>20$~GeV
($\pt>20$~GeV/$c$).  They must fulfill several other identification
criteria designed to select pure samples of high-$\pt$ electrons
(muons)~\cite{LepSel}, including an isolation requirement that the
energy within a cone of $\Delta R < 0.4$ around the lepton axis is
less than 10\% of the $\et$ ($\pt$) of the electron (muon).   The jet energies are corrected for detector
effects~\cite{jet_details}.  We require the highest-$\et$ jet in the
event to have $\et > 25$~GeV and the second highest-$\et$ jet in the
event to have $\et > 20$~GeV.  Finally, we require $\met > $ 20~GeV.

Some criteria are imposed specifically on events collected by the
$\met$ plus jets trigger to ensure a high efficiency.
We require that the two jets are sufficiently separated, $\Delta R > 1$, that one of the jets is
central, $|\eta_{\rm jet}|<0.9$, and that the transverse energy of
both jets is larger than 25~GeV.  Even after these cuts, this trigger
path is not fully efficient, which is taken into account by a correction 
curve as a function of $\met$.

\subsection{Backgrounds}

The baseline selection is based on the signal topology we are trying
to select.  However, several backgrounds can result in events with a similar
topology.
\begin{itemize}

\item $W+$jets: events in which a $W$ boson is produced in association with quarks or gluons form a background if the $W$ boson decays leptonically.  This is the dominant background because of its high production cross section and signal-like properties.

\item $Z+$jets: events in which a $Z$ boson is produced in association with two quarks or gluons may enter our signal sample if the $Z$ boson decays to electrons or muons and one lepton falls outside the fiducial region of the detector or other mismeasurement leads to significant $\met$.

\item QCD non-$W$: events in which several jets are produced, but no real $W$ boson is present, may form a background if a jet fakes an electron or muon and mismeasurement of the jet energies results in incorrectly assigning a large $\met$ to the event. \\

\item $t\bar{t}$: top quark pair production is a background because top quarks nearly always decay to a $W$ boson and a $b$ quark.  If a $W$ boson decays leptonically, $t\bar{t}$ events may pass our baseline event selection criteria.

\item Single top: leading-order production and decay of single top quarks results in an event topology with a $W$ boson and two quarks.

\end{itemize}

\subsection{Event vetoes}

In order to reduce the size of the backgrounds described above,
several vetoes are imposed on events in our sample.  Events are
required to have no additional electrons, muons, or jets, reducing the
$Z+$jets, QCD non-$W$, and $t\bar{t}$ backgrounds.  A further $Z$+jets
veto rejects events with a second loosely identified lepton with the
opposite charge as the tight lepton if the invariant mass of the tight
and loose lepton pair is close to the $Z$ boson mass: $76 < M_{\ell\ell} <
106$~GeV/$c^2$. $ZZ$ events are also effectively removed after this veto.

A veto developed specifically to reduce the size of the QCD non-$W$
background is imposed.  This veto is more stringent for events which
contain an electron candidate, since jets fake electrons more often
than muons.  In electron events, the minimum $\met$ is raised to
25~GeV, and the transverse mass of the leptonically decaying $W$ boson
candidate,
 $M_T(W) =\sqrt{2p_{T,\ell} \MET (1-\cos(\Delta\phi_{\ell,\met}))}$, 
 is required to be at least 20~GeV/$c^2$.  A
variable called the $\met$ significance is also defined:
\begin{widetext}
\begin{equation}
\met^{\rm sig} = \frac{\met}
{\sqrt{ \displaystyle \sum_{\rm jets} C^2_{\rm JES} \cos^2(\Delta \phi_{{\rm jet,}\vec{\not{E}}_T})E_{T, {\rm jet}}^{\rm raw} + \cos^2(\Delta \phi_{\vec{E}_{T,{\rm uncl}},\vec{\not{E}}_T}) \sum E_{T, {\rm uncl}}}},
\end{equation}
\end{widetext}
where $E_T^{\rm raw}$ is the raw, uncorrected energy of a jet and
$C_{\rm JES}$ is the correction to the jet energy~\cite{jet_details},
$\vec{E}_{T,{\rm uncl}}$ is the vector sum of the transverse component
of calorimeter energy deposits not included in any jet, and $\sum
E_{T, {\rm uncl}}$ is the total magnitude of the unclustered
calorimeter energies.  The $\met^{\rm sig}$ is a measure of the
distance between the $\met$ and jets or unclustered energy; it tends to be larger for $\met$ stemming from a neutrino than for $\met$
stemming from mismeasurement.  We require $\met^{\rm
sig}>0.05M_T(W)+3.5$ and $\met^{\rm sig} > 2.5-3.125\Delta\phi_{\met,
jet2}$ in events with an electron candidate.  In muon events, the QCD
veto simply requires $M_T(W) > 10$~GeV/$c^2$.

We veto events with additional ``loose'' jets, defined as jets with $\et >
12$~GeV and $|\eta|<2.0$.  This veto is found to improve the agreement
between Monte Carlo and data in the modeling of some kinematic
variables.

Events consistent with photon conversion and cosmic ray muons are also
vetoed~\cite{stPRD}.

\section{Modeling}
\label{sec:MC}

Both the normalization (number of events in our sample) and 
the shapes of signal and background processes must be
understood to carry out this analysis.

\subsection{Models used}

The signal processes and all background processes except the QCD
non-$W$ background are modeled using events generated by a Monte Carlo
program which are run through the CDF II detector
simulation~\cite{CDFsim}.  The Monte Carlo event generators used for each
process are listed in Table~\ref{tab:MCgen}.  {\sc pythia} is a
leading-order event generator that uses a parton shower to account for
initial and final state radiation~\cite{pythia}.  {\sc alpgen} and {\sc madevent} are 
leading-order parton-level event generators~\cite{alpgen,madevent}; events
generated by {\sc alpgen} and {\sc madevent} are passed to {\sc pythia} where the parton
shower is simulated. 

The top mass is assumed to be $175$~GeV/$c^2$ in the modeling of
$t\bar{t}$ and single top events.  The distributions of the
longitudinal momenta of the different types of quarks and gluons
within the proton as a function of the momentum transfer of the
collision are given by parton distribution functions (PDFs).  The
CTEQ5L PDFs are used in generating all Monte Carlo samples in this
analysis~\cite{CTEQ}.

\begin{table}[ht]
\caption{\label{tab:MCgen}Monte Carlo programs used to generate events for signal and background processes.}
\begin{ruledtabular}
\begin{tabular}{ll}
Sample & Generator \\
\hline
$WW$ & {\sc pythia} \\
$WZ$ & {\sc pythia} \\
$W$+jets & {\sc alpgen} + {\sc pythia} \\
$Z$+jets & {\sc alpgen} + {\sc pythia} \\
$t\bar{t}$ & {\sc pythia} \\
Single top & {\sc madevent} + {\sc pythia} \\
\end{tabular}
\end{ruledtabular}
\end{table}
Simulation of the QCD non-$W$ background is difficult: its
production cross section is large and the probability to mimic a $W$
boson in the event is small.  In addition, the mismeasurements that
lead to the QCD non-$W$ background having large $\met$ may not be
simulated well. Therefore, this background is modeled using data 
rather than simulation.  Events from jet-based triggers containing a
jet that deposits most of its energy in the electromagnetic segment of
the calorimeter, as well as events from single lepton triggers that
fail lepton requirements but pass a looser set of requirements are
used.

\subsection{Expected event yields}
\label{sec:yield}

The number of events due to the signal and $Z+$jets, $t\bar{t}$ and
single top backgrounds that enter our sample are estimated based on
their cross section ($\sigma$), the efficiency ($\epsilon$) with which
they are selected, and the integrated luminosity ($\mathcal L$): $N = \epsilon
{\mathcal L} \sigma$.  The efficiency $\epsilon$, which includes the detector acceptance, is
estimated from the Monte Carlo simulation.  $\sigma$ is taken from NLO
calculations for the $WW$, $WZ$, $t\bar{t}$ and single top processes
and from the CDF inclusive $Z$ boson production cross section measurement for the
$Z$+jets background~\cite{xsections}.  

As mentioned in the introduction, the $WW$ and $WZ$ cross sections
calculated at NLO are $11.66\pm0.70$~pb and $3.46 \pm 0.30$~pb
respectively~\cite{VVtheory}.  The acceptance of these samples
measured with respect to the inclusive production cross section is
about 2.4\% for $WW$ events and about 1.2\% for $WZ$ events.

\begin{figure}[ht]
  \centering \includegraphics[width=0.4\textwidth]{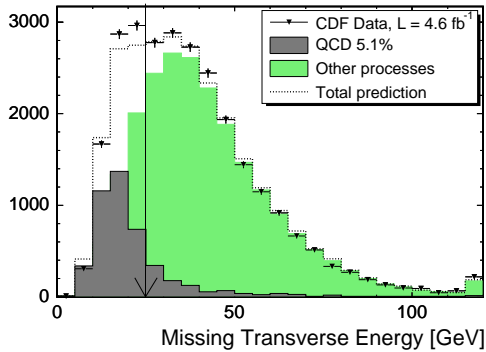}
  \caption{\label{fig:METfit}Fit to $\met$ to determine the contribution from the QCD non-$W$ background for events containing a central electron.}
\end{figure}

Since neither the production cross section nor the selection
efficiency of the QCD non-$W$ background is known, we rely on a
data-driven technique to estimate its normalization.  The shape of the
$\met$ spectrum is very different in events with a real $W$ boson than
in the events coming from the QCD non-$W$ background, as is shown in
Fig.~\ref{fig:METfit}.  The $\met$ spectrum observed in data is fit
with the sum of all contributing processes, where the QCD non-$W$
normalization and the $W+$jets normalization are free parameters.  The
fit is performed over $0<\met<120$ GeV, meaning the cut on the $\met$
described in the event selection above is removed.  An example of the
fit is shown in Fig.~\ref{fig:METfit} for events with a central
electron.  The percentage of QCD non$W$ events in our signal sample (with the
$\met$ cut imposed) is estimated based on the fit; it is about 5\% for
events with a central electron, 3\% for events with a central muon,
and 3\% for events in the extended muon category.

The $W$+jets normalization is a free parameter in the final likelihood fit to 
extract the $WW+WZ$ cross section, which is described in Section VC.  
A preliminary estimate of the $W$+jets normalization used in the modeling validation is derived from
the $\met$ fit described above. Table~\ref{tab:yields} lists the total 
expected number of events for signal and background processes. 
The background normalization uncertainties will be described in Sec.~\ref{sec:Sys}.


\begin{table}[ht]
\caption{\label{tab:yields}Expected number of events for each signal and background process.}
\begin{ruledtabular}
\begin{tabular}{lc}
Process & Predicted number of events \\
\hline
$WW$ signal & 1262 $\pm$ 110 \\
$WZ$ signal & 191 $\pm$ 21 \\
$W$+jets & 35717 $\pm$ 7143 \\
Non-$W$ & 1515 $\pm$ 606 \\
$Z$+jets & 1680 $\pm$ 220 \\
$t\bar{t}$ & 285 $\pm$ 38 \\
Single top & 267 $\pm$ 40 \\
\end{tabular}
\end{ruledtabular}
\end{table}

\subsection{Background shape validation}

\begin{figure*}
\centering \includegraphics[width=0.4\textwidth]{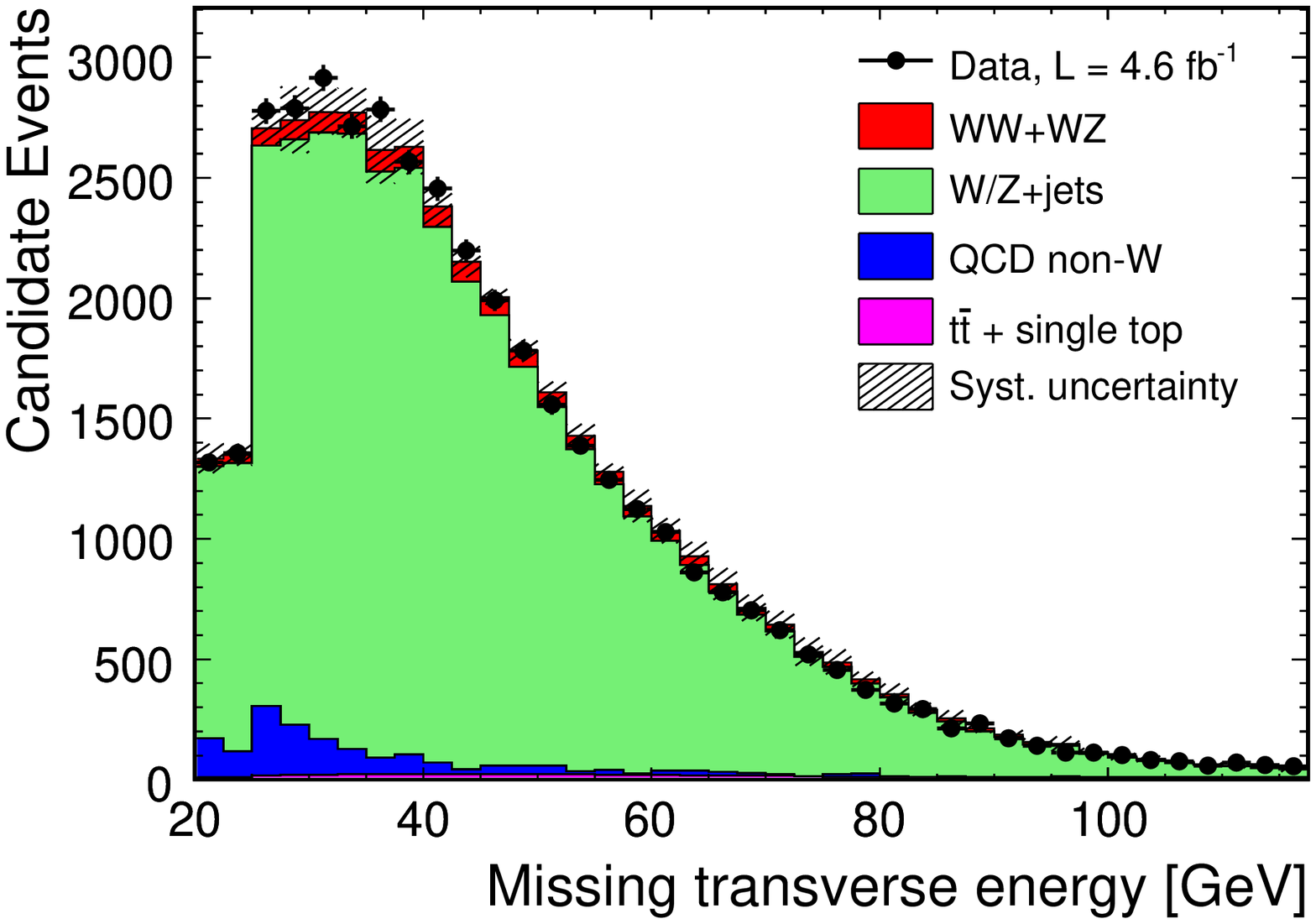}
  \centering \includegraphics[width=0.4\textwidth]{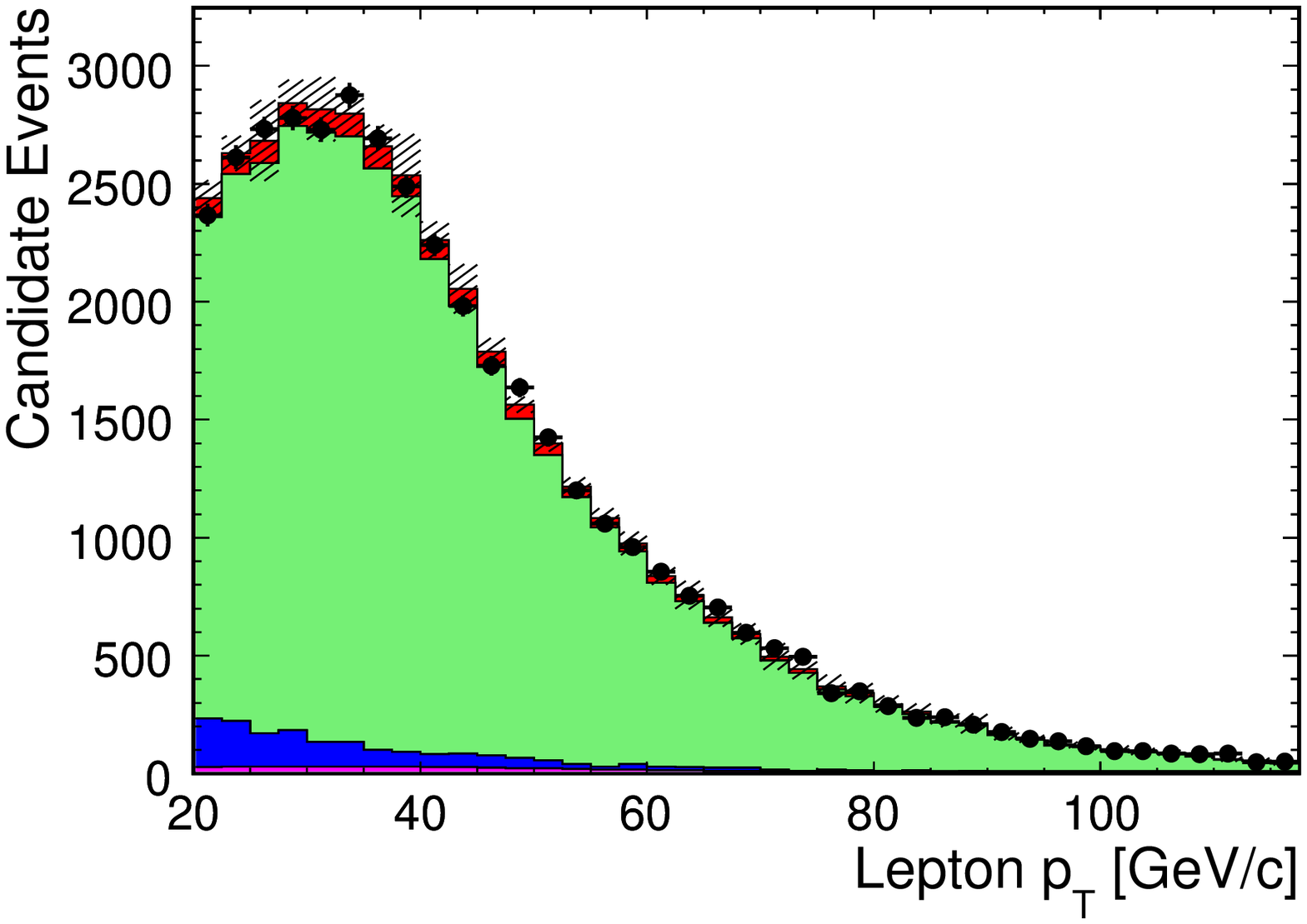}
  \centering \includegraphics[width=0.4\textwidth]{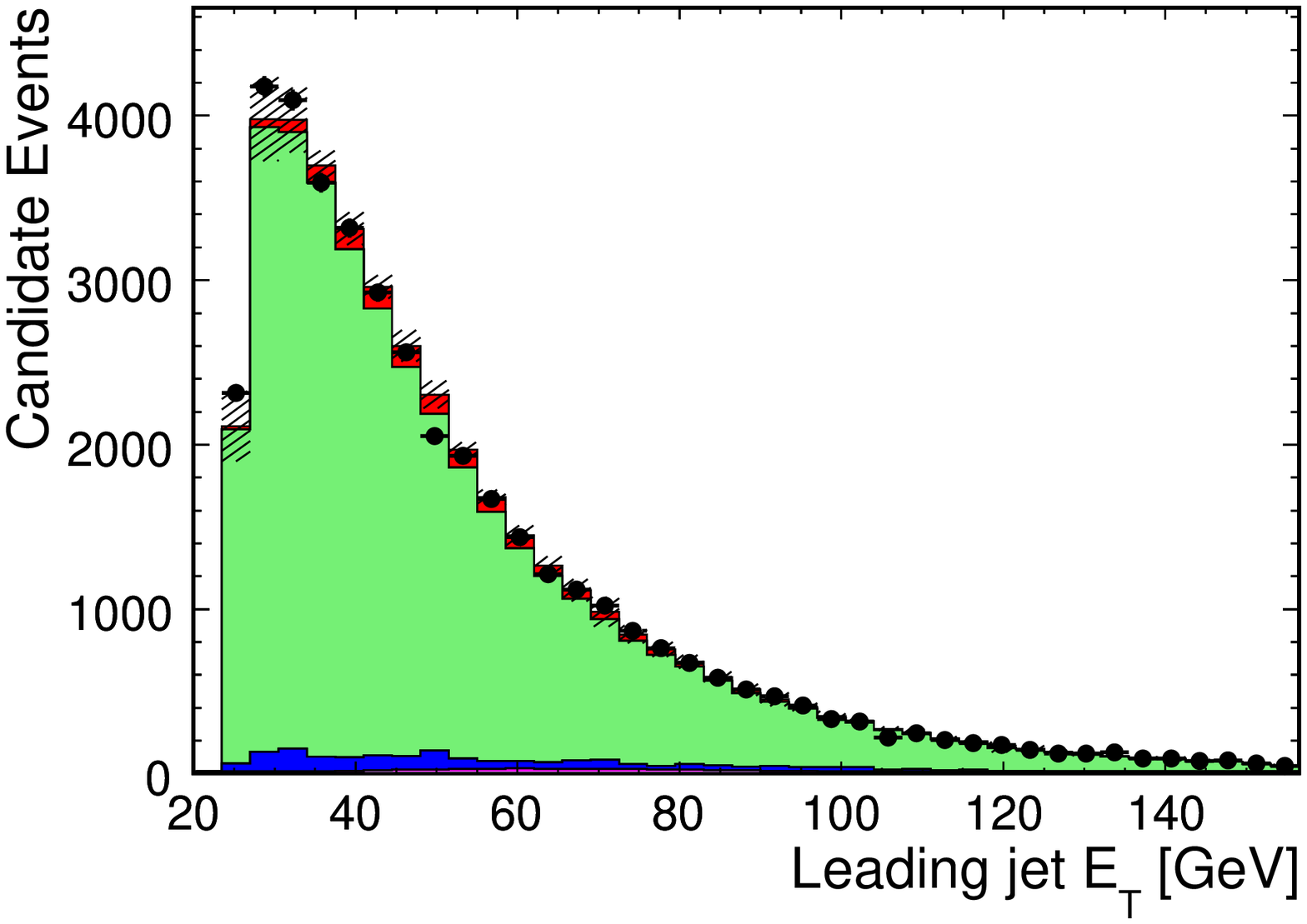}
  \centering \includegraphics[width=0.4\textwidth]{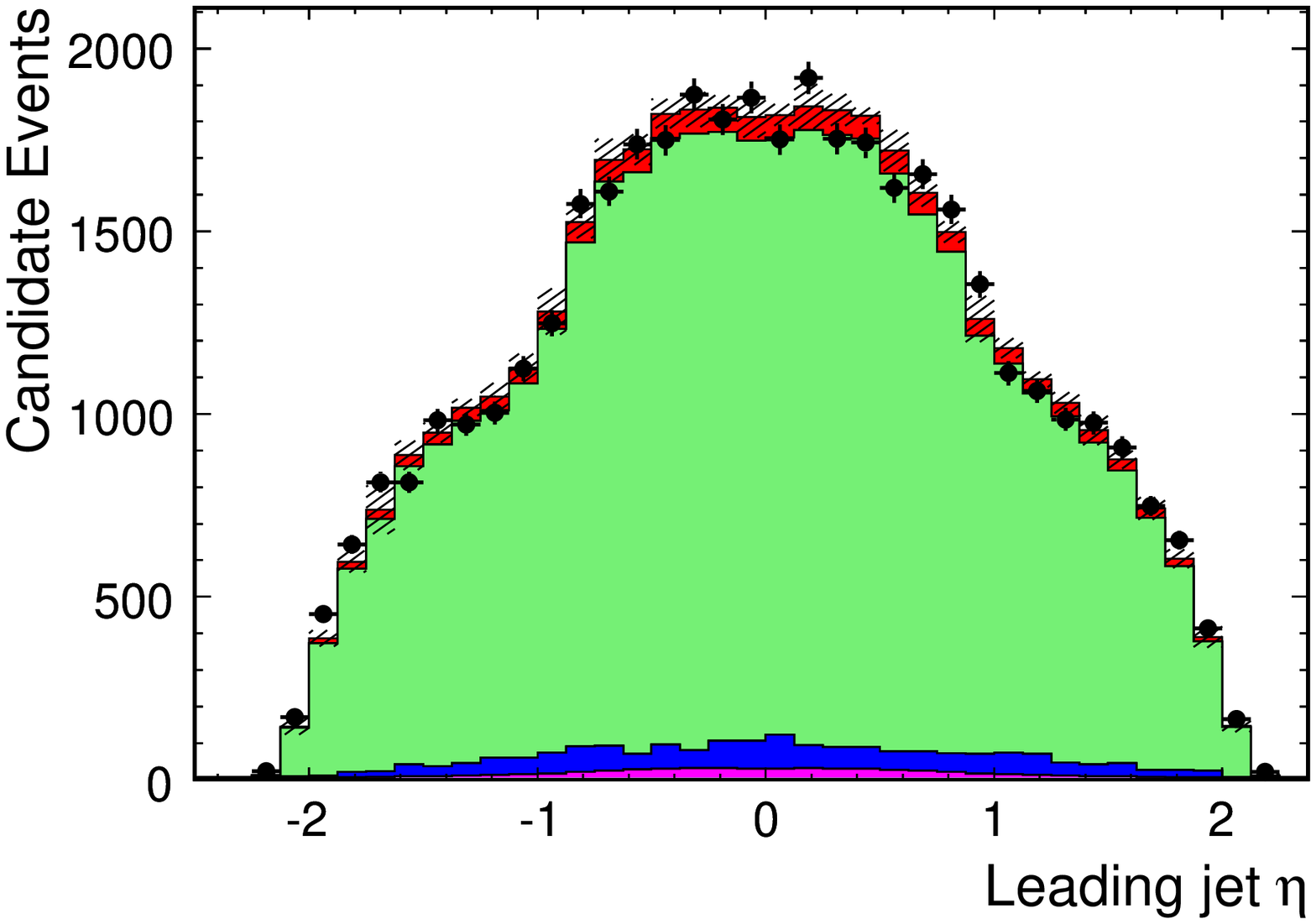}
  \centering \includegraphics[width=0.4\textwidth]{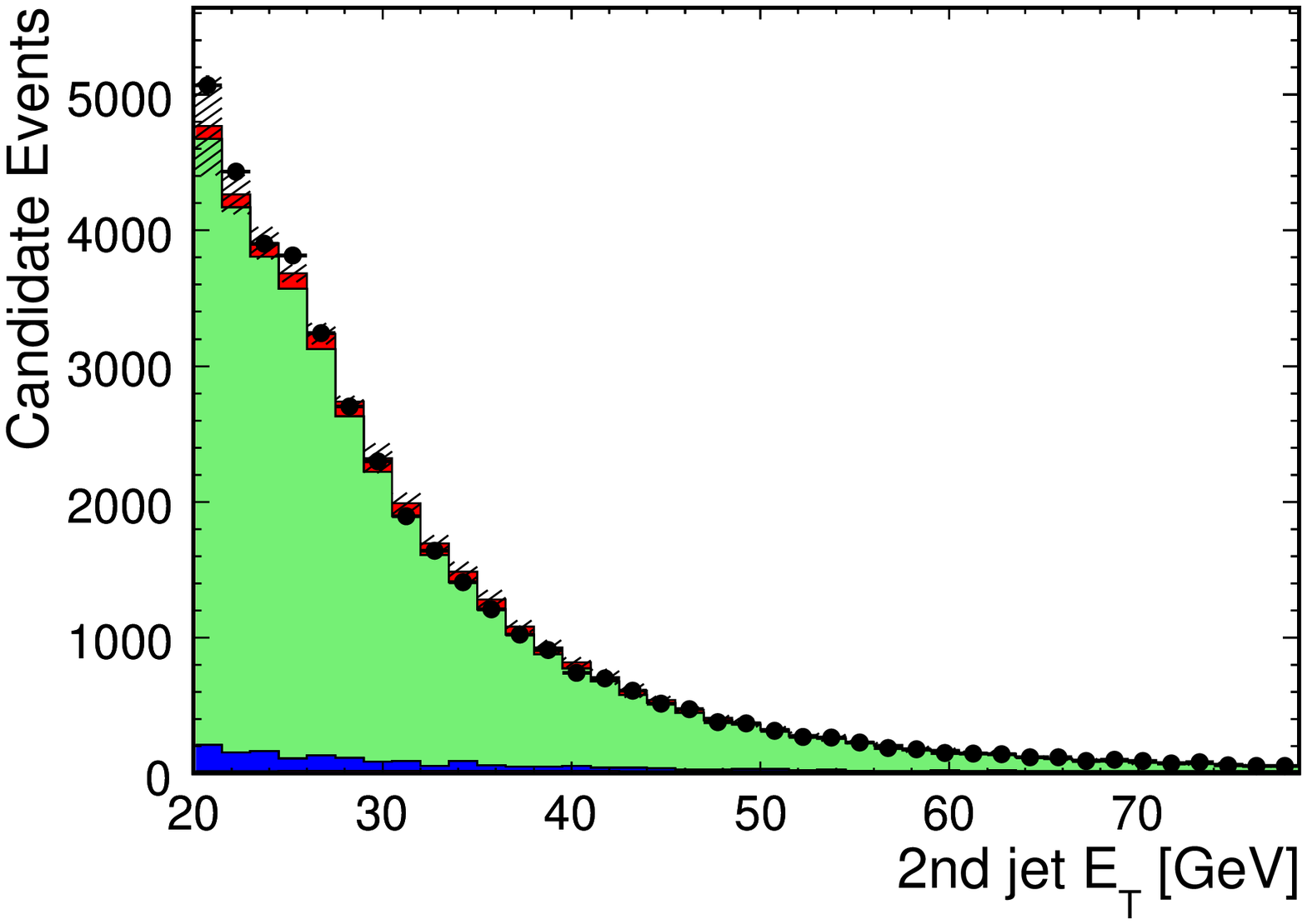}
  \centering \includegraphics[width=0.4\textwidth]{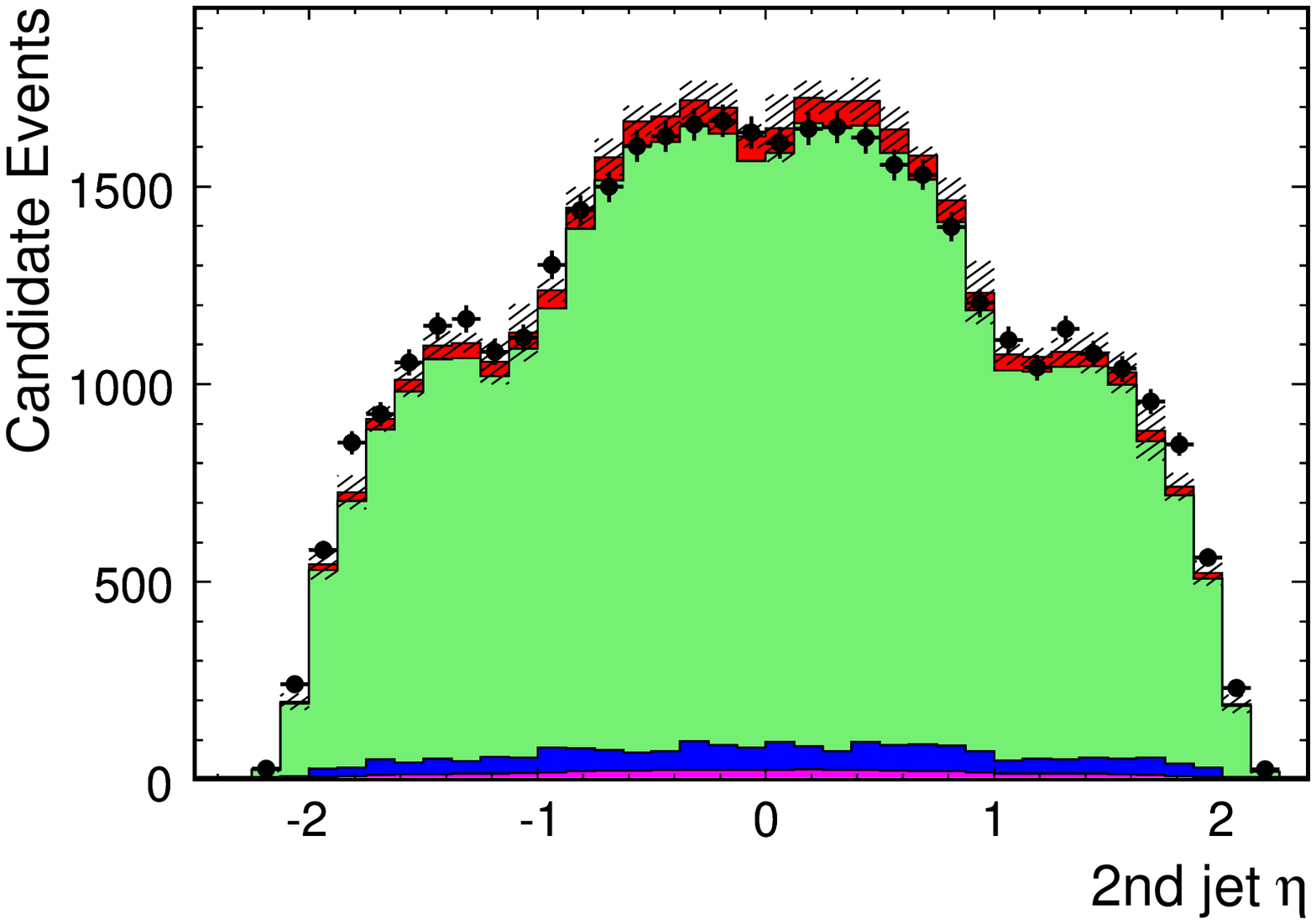}
  \centering \includegraphics[width=0.4\textwidth]{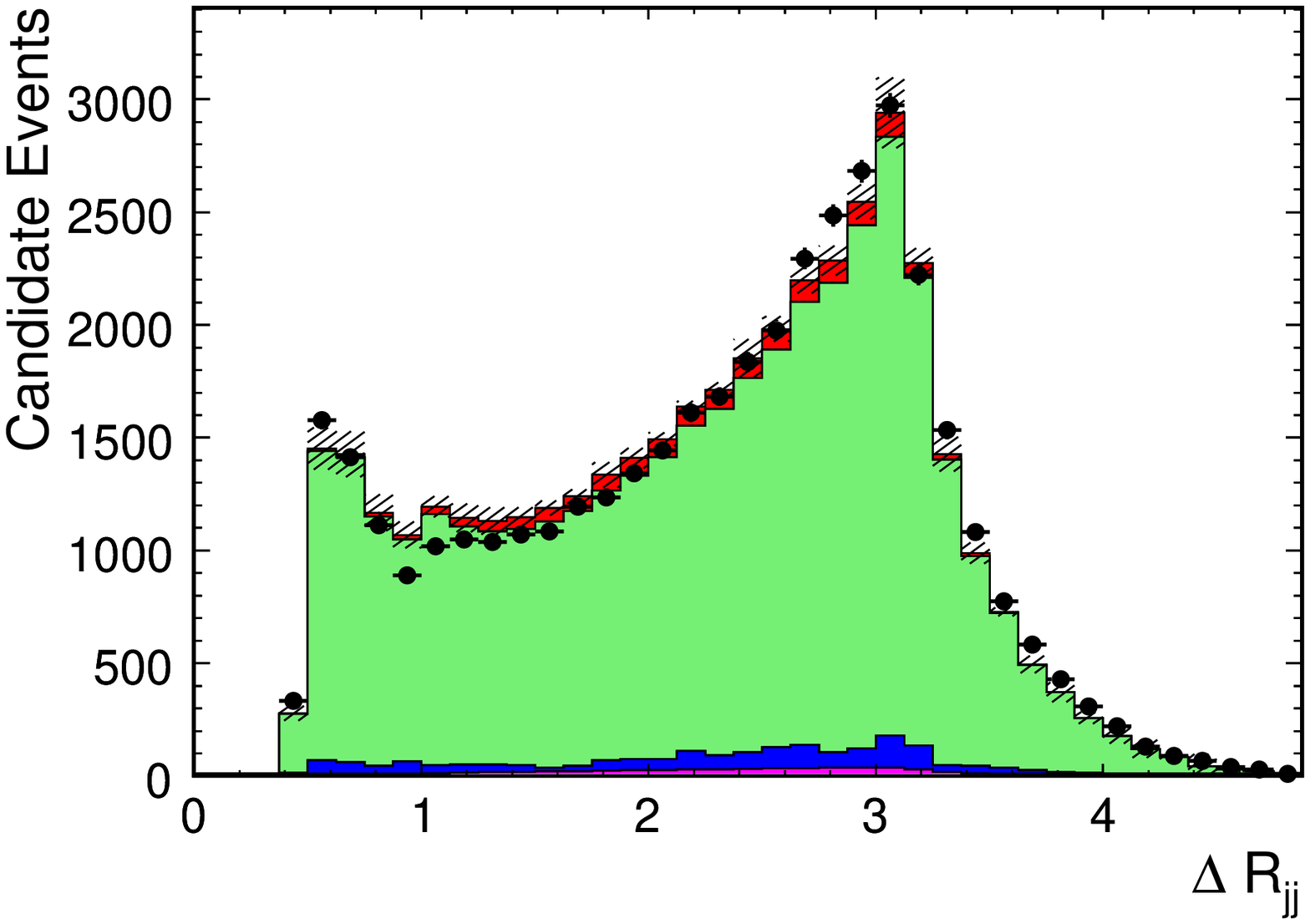}
  \centering \includegraphics[width=0.4\textwidth]{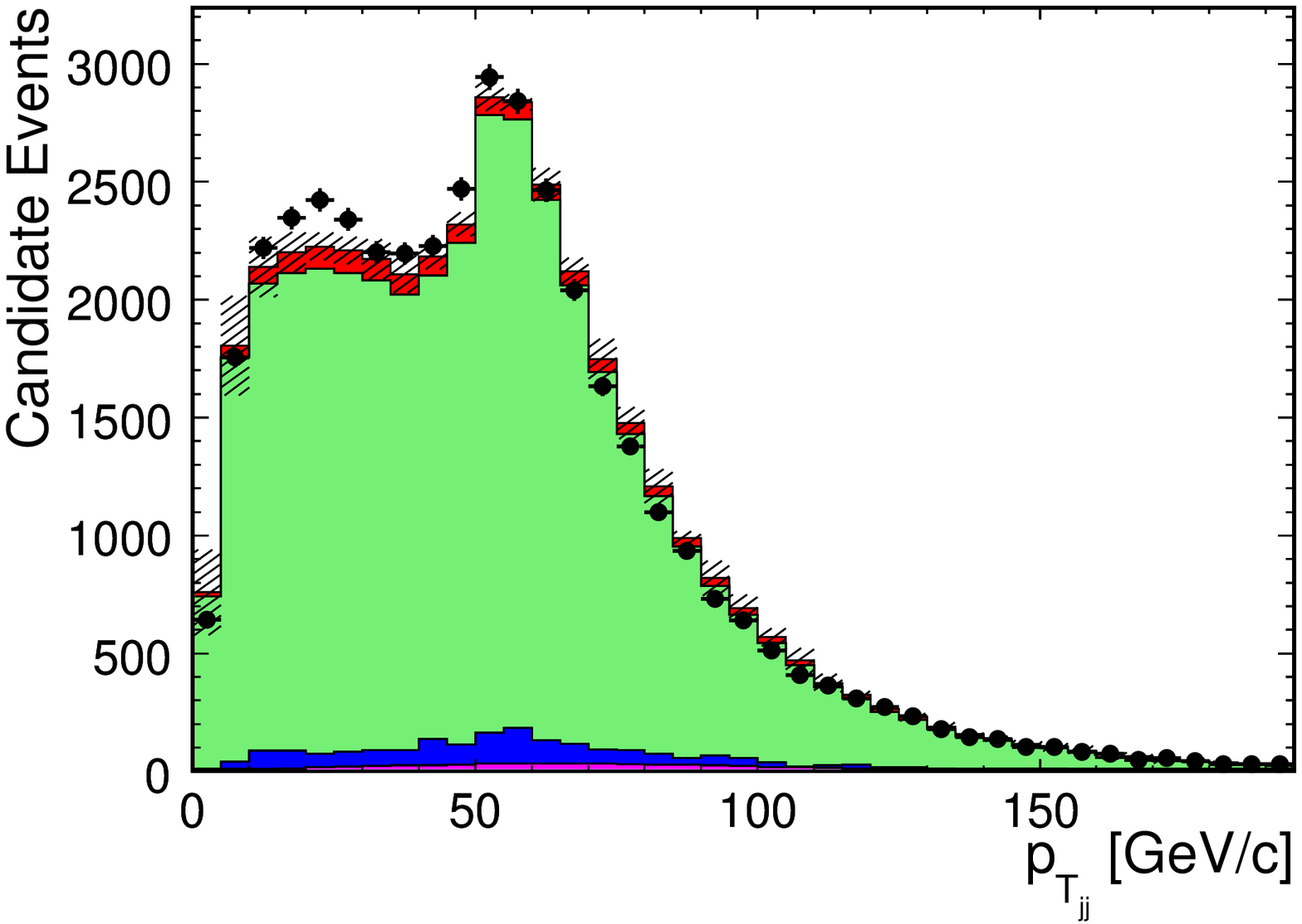}
  \caption{\label{fig:val}Comparison of shapes between data and models for various kinematic quantities. The shaded region includes the effect of the major systematic uncertainties: the jet energy scale, JES, and the renormalization scale, $Q^2$.}
\end{figure*}

The kinematics of the background model are validated by comparing the
shape of various kinematic quantities in data to the prediction from
the models.  Each signal and background process is normalized
according to Table~\ref{tab:yields}, and the sum of their shapes for a
given quantity is compared to that observed in the data.  Some
examples of the comparisons are shown in Fig.~\ref{fig:val} for the
$\met$, the lepton $\et$, the $\et$ and $\eta$ of both jets, the
distance between the two jets ($\Delta R_{jj}$), and the $\pt$ of the two-jet system
($\pt_{jj}$).  In all of these figures, the integral of the total
expectation is set to be equal to the number of data events, so the
figures show shape comparisons. The hatched band is the uncertainty
in the shape of the backgrounds due to the jet energy scale and the
$Q^2$ scale in {\sc alpgen}, described further in Sec.~\ref{sec:Sys}.
The modeling of the kinematic quantities generally matches the data
well within the uncertainties.  In the case of $\pt_{jj}$, the
systematic uncertainties do not seem to cover the disagreement
between data and Monte Carlo, so an additional mismodeling
uncertainty is imposed; this is described further in
Sec.~\ref{sec:Sys}.  The mismodeling uncertainty derived from
$\pt_{jj}$ also affects the modeling of correlated variables,
particularly $\Delta R_{jj}$ and $M_{jj}$, covering the observed
disagreement between data and expectation.

\section{Measurement technique}
\label{sec:ME}

The expected number of events from $WW+WZ$ production is small
compared to the expected number of events from $W$+jets production.
Moreover, the uncertainty on the number of $W$+jets events expected is
large due to uncertainty in the modeling of this process, making it
difficult to separate the $WW+WZ$ signal from the $W$+jets background.
We employ a matrix element analysis technique to improve the signal
and background separation.  Matrix element probabilities for various
processes are calculated which are then combined to form a single
discriminant.

\subsection{Matrix element event probability}

The matrix element method defines a likelihood for an event to be due
to a given production process based on the differential cross section
of the process.  An outline of the procedure is given here, and full
details can be found in Ref.~\cite{pdongthesis}.  

The differential cross section for an $n$-body final state with two
initial state particles with momenta $\vec{q_1}$ and $\vec{q_2}$ and
masses $m_1$ and $m_2$ is
\begin{equation}
d\sigma = \frac{(2\pi)^4 |\mathcal{M}|^2}{4\sqrt{(\vec{q_1} \cdot \vec{q_2})^2-m_1^2 m_2^2}} \times d\Phi_n
\label{eq:dsigma}	
\end{equation}
where $d\Phi_n$ is a phase space factor given by 
\begin{equation}
d\Phi_n = \delta^4 (q_1+q_2- \displaystyle \sum_{i=1}^{n} p_i) \displaystyle \prod_{i=1}^{n} \frac{d^3p_i}{(2\pi)^32E_i}
\label{eq:phase}
\end{equation}
and $E_i$ and $p_i$ are the energies and momenta of the final state
particles~\cite{PDG}.  $\mathcal{M}$ is the matrix element of the
process.

We define a probability density for a given process by
normalizing the differential cross section to the total cross section:
\begin{equation}
P \sim \frac{d\sigma}{\sigma}.
\end{equation}
$P$ is not a true probability, as various approximations are used in
the calculation of the differential cross section: leading-order
matrix elements are used, there are integrations over unmeasured
quantities (described below), and several constants are omitted from
the calculation.

We cannot measure the initial state momenta and the resolution of the 
final state measurements is limited by detector effects.  As a result, we weight
the differential cross section with parton distribution functions (PDFs) for
the proton and integrate over a transfer function encoding the
relationship between the measured quantities $x$ and the parton-level
quantities $y$.  The probability density is then given by
\begin{equation}
P(x) = \frac{1}{\sigma} \int d\sigma(y) dq_1 dq_2 f(q_1)f(q_2)W(y,x),
\end{equation}
where $f(q_1)$ and $f(q_2)$ are the PDFs in terms of the fraction of
the proton momentum ($q_i = E_{qi}/E_{\rm beam}$), and $W(y,x)$ is the
transfer function.  The PDFs are evaluated based on the CTEQ6.1
parameterization~\cite{CTEQ}.  Using Eqs.~\ref{eq:dsigma} and~\ref{eq:phase}
and neglecting the masses and transverse momenta of the initial
partons, the event probability is given by
\begin{equation}
P(x) = \frac{1}{\sigma} \int 2 \pi^4 |\mathcal{M}|^2 \frac {f(y_1)}{|E_{q1}|} 
\frac {f(y_2)}{|E_{q2}|} W(y, x) d\Phi_4dE_{q1}dE_{q2}.
\end{equation}

The squared matrix element, $|\mathcal{M}|^2$, is calculated at tree
level using the {\sc helas} package~\cite{helas}, with the diagrams
for a given process provided by {\sc madgraph}~\cite{madevent}.

In $W(y,x)$, the lepton energy and angle, as well as the jet angles,
are assumed to be measured exactly.  The jet energy transfer function
is derived by comparing parton energies to the fully simulated jet
response in Monte Carlo events.  A double Gaussian parameterization of
the difference between the jet and parton energy is used.  Three
different transfer functions are derived: one for jets originating
from $b$ quarks, one for jets originating from other non-$b$ quarks,
and one for jets originating from gluons.  The appropriate transfer
function is chosen based on the diagram in the matrix element being
evaluated.  The measured missing transverse energy is not used in the
calculation of the event probability; conservation of momentum is used 
to determine the momentum of the neutrino.

After conservation of energy and momentum have been imposed, the
integral to derive the event probability is three-dimensional: the
energies of the quarks and the longitudinal momentum of the neutrino
are integrated over.  The integration is carried out numerically using
an adaptation of the CERNLIB {\sc radmul} routine~\cite{radmul} or the
faster {\sc divonne} integration algorithm implemented in the {\sc
cuba} library~\cite{cuba}.  The results of the two integrators were
checked against each other and found to be compatible.

\subsection{Event Probability Discriminant}

The matrix element event probability is calculated for the signal $WW$
and $WZ$ processes, as well as for single top production and several
contributions to the $W$+jets background: $Wgg$, $Wgq$, $Wbb$, $Wcc$,
and $Wcg$, where $g$, $q$, $b$, and $c$ are gluons, light flavor
quarks, bottom quarks, and charm quarks respectively.

No matrix element calculation is carried out for the $t\bar{t}$,
$Z$+jets, and QCD non-$W$ background processes.  All of these backgrounds
require some additional assumptions, making the matrix element
calculation more difficult and computationally intensive.  For
example, $t\bar{t}$ events become a background if several jets or a
lepton are not detected; incorporating this in the matrix element
calculation requires additional integrations which are computationally
cumbersome.  For the $Z$+jets background process, a lepton either fakes a jet
or escapes detection, two scenarios difficult to describe in the
matrix element calculation.  Finally, the QCD non-$W$ background
would require a large number of leading-order diagrams as well as a
description of quarks or gluons faking leptons.  The $Z$+jets and QCD
backgrounds look very different from the signal (i.e. there will be no
resonance in the dijet mass spectrum) so we expect good discrimination
even without including probabilities explicitly for those background processes.

The probabilities for individual processes described above ($P_i$,
where $i$ runs over the processes) are combined to form a
discriminant, a quantity with a different shape for background-like
events than for signal-like events.  We define the discriminant to be
of the form $P_{\rm signal}/(P_{\rm background}+P_{\rm signal})$ so
that background-like events will have values close to zero and
signal-like events will have values close to unity.  The $P_{\rm
signal}$ and $P_{\rm background}$ are just the sum of individual
probabilities for signal and background processes, but we put in some
additional factors to form the event probability discriminant, or
$EPD$.

First, as noted above, various constants are omitted from the
calculations of $P_i$.  We normalize the $P_i$ relative to each
other by calculating them for each event in large Monte Carlo samples.
We then find the maximal $P_i$ over all Monte Carlo events
corresponding to a given process, $P_i^{\rm max}$.  The normalized
probabilities are then given by $P_i/P_i^{\rm max}$.

In addition, we multiply each $P_i$ by a coefficient, $C_i$.  This
coefficient has the effect of weighting some probabilities more than
others in the discriminant.  These $C_i$ are optimized to achieve the
best expected sensitivity based on the models.  The full $EPD$ is then
given by:
\begin{equation}
EPD = \frac{ \displaystyle \sum_{i=1}^{n_{\rm sig}} \frac {C_i P_i}{P_i^{\rm max}}}
{\displaystyle \sum_{i=1}^{n_{\rm sig}} \frac {C_i P_i}{P_i^{\rm max}} + 
\displaystyle\sum_{j=1}^{n_{\rm BG}} \frac{C_j P_j}{P_j^{\rm max}}},
\end{equation}
where the summation over signal processes runs over $WW$ and $WZ$
($n_{\rm sig} = 2$) and the summation over background processes runs over
$Wgg$, $Wgj$, $Wbb$, $Wcg$, and the single top diagrams ($n_{\rm BG} = 6$) .

\begin{figure}[ht]
  \centering
  \includegraphics[width=0.48\textwidth]{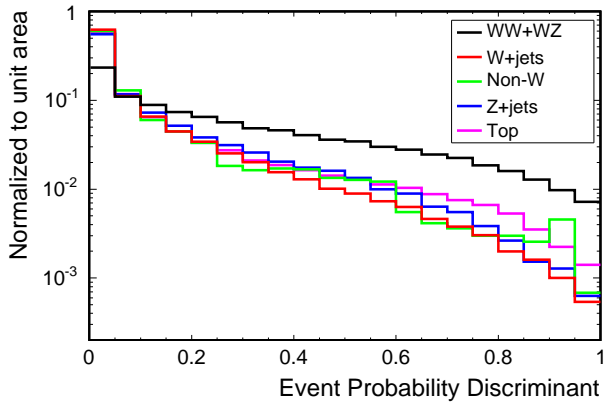}
  \caption{\label{fig:EPD_norm}Shape of the $EPD$ for signal and background processes normalized to unit area.}
\end{figure}

\begin{figure}[ht]
  \centering
  \includegraphics[width=0.48\textwidth]{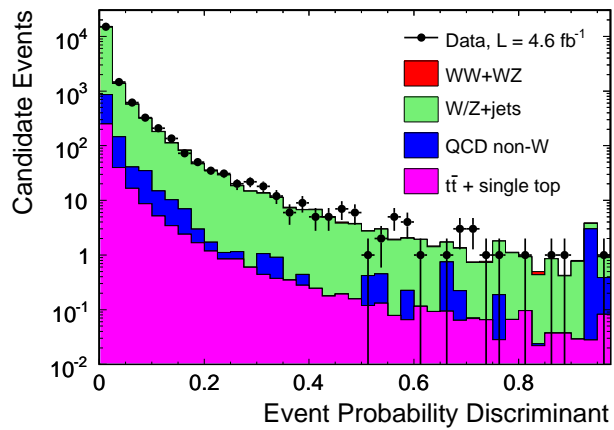}
  \caption{\label{fig:EPD_side}Comparison of the $EPD$ in data and
  simulation for events with $M_{jj}<55$~GeV/$c^2$ or $M_{jj}>120$~GeV/$c^2$.}
\end{figure}

Figure~\ref{fig:EPD_norm} shows the $EPD$ templates for signal and
background processes normalized to unit area.  The background
processes all have similar shapes while the signal process falls more
slowly.  We validate the modeling of the $EPD$ for background events
by comparing data and simulation in the region with
$M_{jj}<55$~GeV/$c^2$ and $M_{jj}>120$~GeV/$c^2$, where we expect very
little signal.  The result of the comparison is shown in
Fig.~\ref{fig:EPD_side}.  The agreement between data and simulation is
very good.

\begin{figure}[ht]
  \centering \includegraphics[width=0.48\textwidth]{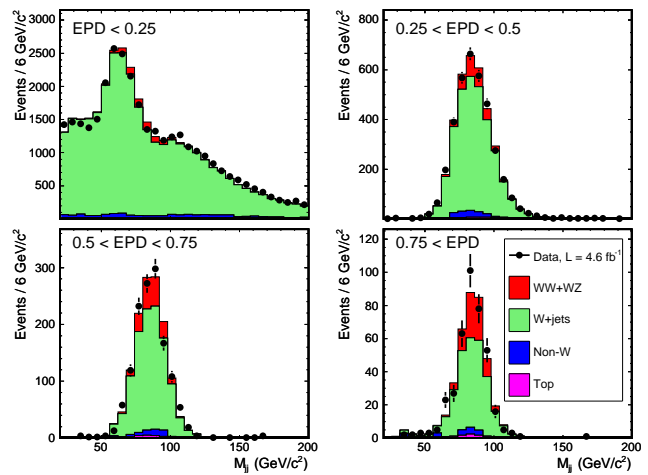}
  \caption{\label{fig:MjjEPDbins}Distribution of the dijet mass in four $EPD$ bins.}
\end{figure}

The effectiveness of the $EPD$ in isolating signal-like events can be
seen by plotting the invariant mass of the two jets in $EPD$ bins,
shown in Fig.~\ref{fig:MjjEPDbins}.  This quantity is expected to have
a resonance around the $W$ or $Z$ boson mass for signal-like events.
The bin with low $EPD$ values (0$<EPD<$0.25), in the top left plot,
has events in the full dijet mass range from 20 to 200~GeV/$c^2$.  For
$EPD>0.25$, however, the $M_{jj}$ distribution is peaked around the
$W/Z$ mass.  As the $EPD$ range approaches unity, the expected signal
to background ratio increases and the dijet mass peak becomes narrower.

\subsection{Likelihood fit}

The shape of the $EPD$ observed in data is fit to a sum of the
templates shown in Fig.~\ref{fig:EPD_norm} to extract the signal
cross section.  The events are divided into three channels
corresponding to different lepton categories: one channel for central
electrons, another for central muons, and a third for events with
muons collected by the $\met$ plus jets trigger.

A maximum likelihood fitting procedure is used.  The likelihood is
defined as the product of Poisson probabilities over all bins of the
$EPD$ template over all channels:
\begin{equation}
L = \displaystyle \prod_{i=1}^{n_{bins}} 
\frac{ {\mu}_i^{n_i}}{n_i!} e^{-{\mu}_i},
\end{equation}
where $n_i$ and $\mu_i$ are the observed and predicted number of
events in bin $i$ respectively.  The prediction in a bin is the sum
over signal and background predictions:
\begin{equation}
{\mu}_i = \displaystyle \sum_{k=1}^{n_{\rm sig}}s_{ik} + \displaystyle
\sum_{k=1}^{n_{\rm bg}}b_{ik}
\label{eq:LikeBin}
\end{equation}
with $b_{ik}$ the predicted contribution from background $k$ in bin
$i$.  $n_{\rm sig}$ is two, corresponding to the $WW$ and $WZ$ processes;
$n_{\rm bg}$ is the number of background processes.

The predicted number of events in a bin is affected by systematic
uncertainties.  The sources of systematic uncertainty are described in
detail in Section~\ref{sec:Sys}.  For each source of uncertainty, a
nuisance parameter is introduced whose value changes the predicted
contribution of a process to a bin.  Each nuisance parameter has a
Gaussian probability density function (p.d.f.) with a mean of zero and
a width given by the 1$\sigma$ uncertainty.  A detailed mathematical
description of the way the nuisance parameters are incorporated in the
likelihood is given in Ref.~\cite{stPRD}.

Finally, with a likelihood that is a function of the observed data,
the signal cross section, the predicted signal and background
contributions, and systematic uncertainties and their corresponding
nuisance parameters, we extract the cross section.  A Bayesian
marginalization technique integrates over the nuisance parameters,
resulting in a posterior probability density which is a function of
the signal cross section.  The measured cross section corresponds to
the maximum point of the posterior probability density, and the 68\% confidence interval
is the shortest interval containing 68\% of the area of the posterior probability density.

The measured cross section is the total cross section of the signal,
$\sigma_{WW+WZ}$.  Assuming the ratio between the $WW$ and $WZ$ cross
sections follows the NLO prediction,
$\sigma_{WW+WZ}=\sigma_{WW}+\sigma_{WZ}$.  If the ratio between the
cross sections is different than the NLO prediction, we are measuring
the total cross section
$\sigma_{WW+WZ}=1.13\sigma_{WW}'+0.56\sigma_{WZ}'$.  Here
$\sigma_{WW}'$ and $\sigma_{WZ}'$ are not assumed to follow NLO predictions.  
The ratio between the $WW$ and $WZ$ acceptances is predicted from 
the signal simulations described in Sec.~IVA.

\section{Systematic Uncertainties}
\label{sec:Sys}

Systematic uncertainties affect the normalization of background
processes, the signal acceptance, and the shape of the $EPD$ for both
background and signal processes.  The sources of systematic
uncertainty and the aspects of the measurement affected by each are
briefly described in this section.  Finally, the expected
contribution of the uncertainties to the $WW+WZ$ cross section
measurement are explored.

\subsection{Sources of uncertainty}

\begin{itemize}
\item {\bf Normalization of background processes:} The uncertainties in the normalization of the background processes are summarized in Table~\ref{tab:BGNormSys}.  The uncertainty on the $W$+jets normalization is taken to be an arbitrarily large number; the fit to extract the cross section constrains the $W$+jets normalization to a few percent, so taking a 20\% uncertainty is equivalent to allowing the $W$+jets normalization to float.  The uncertainty on the $Z+$jets, $t\bar{t}$, and single top backgrounds are derived from the uncertainty in their cross sections and uncertainties on the efficiency estimate.  The 40\% uncertainty on the QCD non-$W$ contribution is a conservative estimate based on differences observed between different choice of sample models.

\begin{table}[ht]
\caption{\label{tab:BGNormSys}Uncertainties in the background normalizations.}
\begin{ruledtabular}
\begin{tabular}{lc}
Background & Normalization uncertainty \\
\hline
$W$+jets & 20\% \\
$Z$+jets & 15\% \\
QCD non-$W$ & 40\% \\
$t\bar{t}$ and single top & 12\% \\
\end{tabular}
\end{ruledtabular}
\end{table}

\item {\bf Jet Energy Scale (JES):} As mentioned above, jet energies are corrected for detector effects.
The corrections have systematic uncertainties associated with
them~\cite{jet_details}.  The size of the 1$\sigma$ uncertainty
depends on the $\et$ of the jet, ranging from about 3\% for jet $\et\sim$ 80 GeV to about 7\% for jet $\et\sim$20 GeV.

The effect of the JES uncertainty on the measurement is estimated by
creating two shifted Monte Carlo samples: one in which the energy of
each jet in each event of our Monte Carlo samples is shifted by
$+1\sigma$ and the second in which each jet energy is shifted by
$-1\sigma$, taking the $\et$-dependence of the uncertainty into
account.  The whole analysis is repeated with the shifted Monte Carlo
samples, including the calculation of the matrix elements.

The JES uncertainty has a small effect on the estimated signal
acceptance because the efficiency of the jet $\et$ selection depends
on the JES.  The size of the acceptance uncertainty is about 1\%.  In
addition, the shape of the $EPD$ templates for the signal processes
and for the dominant $W$+jets background process are affected by the
JES uncertainty.  The change in the background shape is relatively
small compared to the change in the signal shape.  The signal
normalization uncertainty, the signal shape uncertainty, and the
background shape uncertainty are incorporated as a correlated
uncertainty in the likelihood fit.

\item {\bf $Q^2$ scale in {\sc alpgen}:} The factorization and renormalization scale, or $Q^2$ scale, is a
parameter in the perturbative expansion used to calculate matrix
elements in {\sc alpgen}.  Higher-order
calculations become less dependent on the choice of scale, but {\sc alpgen}
is a leading-order generator and its modeling is affected by the
choice of scale. The scale used in generating our central $W$+jets samples is $Q^2 = m_{W}^{2} + \Sigma
m_{T}^{2}$, where $m_W$ is the mass of the $W$ boson, $m_T$ is the
transverse mass, and the summation is over all final-state partons.
{\sc alpgen} $W$+jets samples were generated with this central scale
doubled and divided by two.  These are taken as $\pm 1 \sigma$
uncertainties on the shape of the $W$+jets $EPD$ template.

\item {\bf Integrated luminosity:} The integrated luminosity is calculated based on the $p\bar{p}$ inelastic cross section and the acceptance of CDF's luminosity monitor~\cite{Lumi}.  There is a 6\% uncertainty on the calculation, which is included as a correlated uncertainty on the normalization of all processes except the non-$W$ QCD background and the $W$+jets background, whose normalizations are determined from fits to the data.

\item {\bf Initial and final state radiation:} Comparison between samples simulated with {\sc pythia} and Drell-Yan data, where no FSR is expected, are used to determine reasonable uncertainties for the parameters used to tune the initial and final state radiation in {\sc pythia}~\cite{MtopTemplate}.  The signal
$WW$ and $WZ$ samples were generated with the level of ISR and FSR
increased and decreased, and the change in the acceptance was
estimated.  This results in an uncertainty of
about 5\% on the signal acceptance.

\item {\bf PDFs:} The PDFs used in generating the Monte Carlo samples have some uncertainty associated with them.  The uncertainty on the signal acceptance is estimated in the same way as in Ref.~\cite{MtopTemplate}.  The uncertainty in the signal acceptance is found to be 2.5\%.

\item {\bf Jet Energy Resolution (JER):} A comparison between data and simulation is used to assign an uncertainty on the jet energy resolution~\cite{TopWidth}.  For a jet with measured $\et$ of 40~GeV, the jet energy resolution is $(13 \pm 7$)\%.  The matrix element calculations are repeated for the signal Monte Carlo sample with a higher jet energy resolution, and no change in the shape of the $EPD$ is observed.  A small ($\sim 1\%$) uncertainty on the signal acceptance is assigned.

\item {\bf $W$+jets modeling:} In addition to the shape uncertainties on the $W+$jets $EPD$ due to the JES and $Q^2$ scale, we impose shape uncertainties due to mismodeling of the $\pt$ of the dijet system ($\pt_{jj}$) and the $\eta$ of the lower-$\pt$ jet in the event ($\eta_{j2}$).  We derive the uncertainty due to the mismodeling of these variables by reweighting the $W+$jets Monte Carlo model to agree with data as a
function of either $\pt_{jj}$ or $\eta_{j2}$.  When deriving the
weights, we remove events with $55<M_{jj}<120$ GeV/$c^2$ (the region
in which we expect most of the signal) to avoid biasing the measurement
towards the expected result.  The mismodeling of $\eta_{j2}$ has a
negligible effect on the shape of the $EPD$, whereas the mismodeling
of $\pt_{jj}$ has a small effect on its shape.

\item {\bf Lepton identification efficiency:} There is a 2\% uncertainty on the efficiency with which we can identify and trigger on leptons.  This uncertainty is assigned in the same way as the uncertainty on the integrated luminosity.

\end{itemize}

\subsection{Effect on cross section fit}

Pseudoexperiments are carried out to determine the expected
uncertainty on the $WW+WZ$ cross section.  The pseudoexperiments are
generated by varying the bin contents of each template histogram
according to Poisson distributions as well as randomly setting a value
for each nuisance parameter according to its p.d.f. The likelihood
fit is applied to each pseudoexperiment to extract the $WW+WZ$
cross section.

 In order to estimate the effect of
certain systematic uncertainties, they are taken out of the
pseudo-experiments one-by-one.  The expected statistical uncertainty
(including the uncertainty on the background normalizations) was found
to be 14\% while the total systematic uncertainty is expected to be
16\%.  The total (systematic plus statistical) uncertainty expected on the
$WW+WZ$ cross section is 21\%.  The largest predicted systematic uncertainties are the JES,
$Q^2$ scale, and luminosity uncertainties, which contribute 8\%, 7\%,
and 6\% respectively to the total $\sigma_{WW+WZ}$ uncertainty.

Based on the pseudoexperiments, we can also understand which nuisance
parameters are constrained in the likelihood fit.  The $W$+jets
normalization uncertainty, which has a width of 20\% in the prior
p.d.f., is constrained on average to 1.8\% in the pseudoexperiments.
The first few bins of the $EPD$, which are dominated by the $W$+jets
contribution, establish this constraint, and the effect of the
constraint is to reduce the uncertainty in the $W$+jets normalization
in the high-$EPD$ bins, which are most important to the signal
extraction.

\section{Results}
\label{sec:results}

\begin{figure}[ht]
  \centering
  \includegraphics[width=0.45\textwidth]{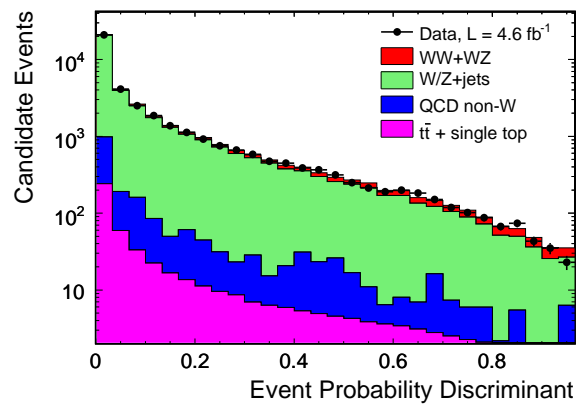}
  \caption{\label{fig:EPDdata}Stacked $EPD$ templates with data
  superimposed.}
\end{figure}

The likelihood fit is carried out in a data sample corresponding to an
integrated luminosity of 4.6~fb$^{-1}$.  The shape of the $EPD$
observed in data is shown superimposed on the shape expected from
Monte Carlo in Fig.~\ref{fig:EPDdata}.  The cross section for
$WW+WZ$ production is found to be $\sigma(p\bar{p} \rightarrow WW+WZ)
= 17.4\pm3.3$~pb.  This result agrees with the prediction from
NLO calculations of $15.1 \pm 0.9$~pb.

The cross section was extracted in each lepton channel separately as a
cross-check.  The results are listed in Table~\ref{tab:results}.  The
extracted cross section agrees across lepton channels.

\begin{table}[ht]
\caption{\label{tab:results}Fitted $WW+WZ$ cross section in the three lepton categories and in the whole sample. }
\begin{ruledtabular}
\begin{tabular}{lc}
Category & Cross section [pb] \\
\hline
Central electrons & $16.3^{+5.1}_{-3.9}$ \\
Central muons & $19.8^{+3.3}_{-5.4}$ \\
Extended muons & $10.7^{+8.6}_{-5.4}$ \\
\hline
All & $17.4\pm3.3$ \\
\end{tabular}
\end{ruledtabular}
\end{table}

\section{Fit to the dijet invariant mass}
\label{sec:Mjj}

\begin{figure}[ht]
  \centering
  \includegraphics[width=0.48\textwidth]{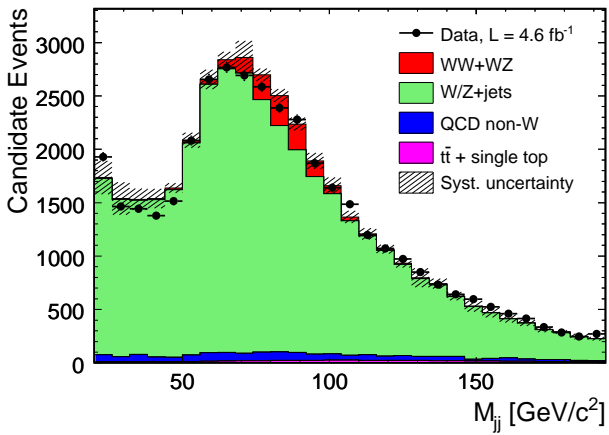}
  \caption{\label{fig:Mjj_val}Distribution of $M_{jj}$ in data superimposed on Monte Carlo prediction. The shaded regions include JES, $Q^2$, and $W$+jets modeling systematic uncertainties.}
\end{figure}

\begin{figure}[ht]
  \centering 
  \includegraphics[width=0.48\textwidth]{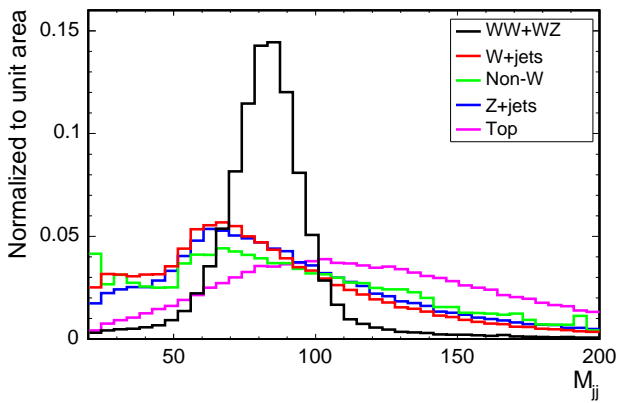}
  \caption{\label{fig:Mjj_norm}Shape of $M_{jj}$ templates for signal and background processes normalized to unit area.}
\end{figure}

A similar template fit to the one described above was carried out
using the invariant mass of the two jets rather than the $EPD$ with
exactly the same event selection and sources of systematic
uncertainty.  The distribution of $M_{jj}$ in data is shown
superimposed on the stacked predictions in Fig.~\ref{fig:Mjj_val}.
The templates for the fit are shown in Fig.~\ref{fig:Mjj_norm}.  There
is a resonance for the $WW+WZ$ signal since the two jets are a product
of $W$ or $Z$ boson decay, while the backgrounds have very different
shapes without an apparent resonance.  The shape of the $W/Z+$jets
background is a falling distribution shaped by event selection cuts.

The expected uncertainty on the $WW+WZ$ cross section extracted by a
fit to $M_{jj}$ is about 19\%, lower than the expected uncertainty
when fitting the $EPD$.  While the statistical uncertainty is larger
when fitting $M_{jj}$ than when fitting the $EPD$, the systematic
uncertainty is smaller.  The dominant systematic uncertainty is
expected to be the shape uncertainty on the $W$+jets background due to
the mismodeling of $\pt_{jj}$, while the JES and $Q^2$ scale
uncertainties are less important than when fitting the $EPD$.

The $WW+WZ$ cross section extracted from the fit to $M_{jj}$ is
$12.4^{+2.7}_{-3.0}$~pb.  Based on pseudo-experiments, the expected
correlation between the fit to $M_{jj}$ and the fit to $EPD$ is about
60\%.  Thus the cross sections extracted from the $EPD$ and the
$M_{jj}$ fits have a discrepancy of about 1.8$\sigma$.

Fitting the dijet mass is presented here as a cross-check to the
result from the matrix element technique because it is a less
sensitive way of extracting the signal.  In other words, the expected
probability that the signal can be faked by the background is higher
when fitting the dijet mass than when fitting the $EPD$.  As a result,
the first observation of the $WW+WZ$ signal in this channel was
provided by the matrix element technique~\cite{ourPRL}.
With the data sample presented in this paper, the expected sensitivity
of the matrix element technique is 5.0$\sigma$, while it is
4.6$\sigma$ when fitting $M_{jj}$.  The observed significances are
5.4$\sigma$ and 3.5$\sigma$ for the matrix element and $M_{jj}$
analyses respectively.

\section{Conclusions}
\label{sec:conc}

We have extracted the cross section for $WW+WZ$ production in the
final state with a lepton, two jets, and missing transverse energy
using a matrix element technique.  The cross section is measured to be
$17.4\pm3.3$~pb, in agreement with the NLO theoretical
prediction of $15.1\pm0.9$~pb.  The measurement is primarily 
systematically limited; the jet energy scale and $Q^2$ scale
uncertainties give both large contributions to the total uncertainty.
Improvements to the cross section measurement could be achieved by 
reducing the size of the systematic uncertainties via data-driven methods.  
The effect of systematic uncertainties on the measurement could also 
be reduced by further optimization of the event selection and discriminant.

\begin{acknowledgments}
We thank the Fermilab staff and the technical staffs of the
participating institutions for their vital contributions. This work
was supported by the U.S. Department of Energy and National Science
Foundation; the Italian Istituto Nazionale di Fisica Nucleare; the
Ministry of Education, Culture, Sports, Science and Technology of
Japan; the Natural Sciences and Engineering Research Council of
Canada; the Humboldt Foundation, the National Science Council of the
Republic of China; the Swiss National Science Foundation; the
A.P. Sloan Foundation; the Bundesministerium f\"ur Bildung und
Forschung, Germany; the Korean Science and Engineering Foundation and
the Korean Research Foundation; the Science and Technology Facilities
Council and the Royal Society, UK; the Institut National de Physique
Nucleaire et Physique des Particules/CNRS; the Russian Foundation for
Basic Research; the Ministerio de Ciencia e Innovaci\'{o}n, and
Programa Consolider-Ingenio 2010, Spain; the Slovak R\&D Agency; and
the Academy of Finland.
\end{acknowledgments}

\pagebreak


\begin{thebibliography}{30}
\bibitem{hagiwara} K. Hagiwara, S. Ishihara, R. Szalapski, and D. Zeppenfeld, Phys. Rev. D {\bf 48}, 2182 (1993).

\bibitem{VVtheory} J.~M.~Campbell and R.~K.~Ellis, Phys.\ Rev.\ D {\bf 60}, 113006 (1999).


\bibitem{diblepCDF} T. Aaltonen {\it et al.} (CDF Collaboration), Phys. Rev. Lett. {\bf 104}, 201801 (2010); 
A. Abulencia {\it et al.} (CDF Collaboration), {\it ibid} {\bf 98}, 161801 (2007); 

\bibitem{diblepD0} V.~Abazov {\it et al.} (D0 Collaboration), Phys. Rev. Lett. {\bf 103}, 191801 (2009) and Phys. Rev. D {\bf 76}, 111104(R) (2007).

\bibitem{metjets} T.~Aaltonen {\it et al.} (CDF Collaboration), Phys.\ Rev.\ Lett.\ {\bf 103}, 091803 (2009).
\bibitem{d0lvjj} V.~M.~Abazov {\it et al.} (D0 Collaboration), Phys.\ Rev.\ Lett.\ {\bf 102}, 161801 (2009).
\bibitem{ourPRL} T.~Aaltonen {\it et al.} (CDF Collaboration), Phys.\ Rev.\ Lett.\ {bf 104}, 101801 (2010).

\bibitem{WHME}  T.~Aaltonen {\it et al.}  (CDF Collaboration),
 Phys.\ Rev.\ Lett.\  {\bf 103}, 101802 (2009).

\bibitem{CDFdet} D. Acosta {\it et al.} (CDF Collaboration), Phys. Rev. D {\bf 71}, 032001 (2005).
\bibitem{SVX} A. Sill {\it et al.}, Nucl.\ Instrum.\ Methods A {\bf 447}, 1 (2000).
\bibitem{COT} T. Affolder {\it et al.}, Nucl.\ Instrum.\ Methods A {\bf 526}, 249 (2004).
\bibitem{CEM} L. Balka {\it et al.}, Nucl.\ Instrum.\ Methods A {\bf 267}, 272 (1988).
\bibitem{CHAWHA} S. Bertolucci {\it et al.}, Nucl.\ Instrum.\ Methods A {\bf 267}, 301 (1988).
\bibitem{PEM} M. Albrow {\it et al.}, Nucl.\ Instrum.\ Methods A {\bf 480}, 524 (2002).
\bibitem{ShowerMax} G. Apollinari {\it et al.}, Nucl.\ Instrum.\ Methods A {\bf 412}, 515 (1998).
\bibitem{CMU} G. Ascoli {\it et al.}, Nucl.\ Instrum.\ Methods A {\bf 268}, 33 (1988).
\bibitem{CLC} D. Acosta {\it et al.}, Nucl.\ Instrum.\ Methods A {\bf 494}, 57 (2002). 
\bibitem{XFT} E.J. Thomson {\it et al.}, IEEE Trans.\ on Nucl.\ Science. {\bf 49}, 1063 (2002).

\bibitem{LepSel} A. Abulencia {\it et al.}, J. Phys.G {\bf 34}, 2457 (2007).
\bibitem{jet_details} A. Bhatti {\it et al.}, Nucl.\ Instrum.\ Methods A {\bf 566}, 375 (2006).

\bibitem{stPRD} T.~Aaltonen {\it et al.} (CDF Collaboration), arXiv:hep-ex/1004.1181

\bibitem{CDFsim} E. Gerchtein and M. Paulini, CHEP03 Conference Proceedings, 2003.

\bibitem{pythia} T.~Sj\"ostrand {\it et al.}, Comput.\ Phys.\ Commun., {\bf 135},\ 238 (2001).
\bibitem{alpgen} M.~L.~Mangano {\it et al.}, J.\ High\ Energy\ Phys.\ 07 (2003) 001.
\bibitem{madevent} J. Alwall {\it et al.} J.\ High\ Energy\ Phys.\ 09 (2007) 028.
\bibitem{CTEQ} J.~Pumplin {\it et al.} J.\ High\  Energy\ Phys.\ 07 (2002) 012.

\bibitem{xsections} D.~Acosta {\it et al.} (CDF collaboration), Phys. Rev. Lett. {\bf 94}, 091803 (2005); M.~Cacciari {\it et al.}, J.\ High\ Energy\ Phys.\ 09 (2008) 127; B.~W.~Harris {\it et al.}, Phys.\ Rev.\  D {\bf 66}, 054024 (2002).

\bibitem{pdongthesis} P. J. Dong, Ph.D. Thesis, University of California at Los Angeles, 2008, FERMILAB-THESIS-2008-12.
\bibitem{PDG} C. Amsler {\it et al.} Phys.\ Lett.\ B {\bf 667}, 1 (2008).
\bibitem{helas} I. Murayama, H. Watanabe and K. Hagiwara, Tech. Rep. 91-11, KEK (1992).
\bibitem{radmul} A.~Genz and A.~Malik, J. Comput. Appl. Math. {\bf 6}, 295 (1980); implemented as CERNLIB algorithm D120, documented at http://wwwasdoc.web.cern.ch/wwwasdoc/shortwrupsdir/d120/top.html.

\bibitem{cuba} T.~Hahn, Comput. Phys. Commun. {\bf 168}, 78 (2005).

\bibitem{Lumi} D. Acosta {\it et al.} Nucl.\ Instrum.\ Methods A {\bf 494}, 57 (2002).
\bibitem{MtopTemplate} A. Abulencia {\it et al.} (CDF Collaboration), Phys. Rev. D {\bf 73}, 032003 (2006).
\bibitem{TopWidth} T. Aaltonen {\it et al.} (CDF Collaboration), Phys. Rev. Lett. {\bf 102}, 042001 (2009).

\end{thebibliography}
\end{document}